\documentclass[openany]{amsbook}
\usepackage[utf8]{inputenc}
\usepackage{titlesec}
\usepackage[colorlinks = true,
            linkcolor = blue,
            urlcolor  = blue,
            citecolor = blue,
            anchorcolor = blue]{hyperref}
\usepackage{float}
\usepackage{placeins}
\usepackage{wrapfig}
\usepackage{listings}
\usepackage{fix-cm}
\usepackage{xcolor}
\usepackage{graphicx}
\usepackage{caption}
\usepackage{subfigure}

\titleformat*{\section}{\Large\bfseries}
\titleformat*{\subsection}{\Large\bfseries}
\titleformat*{\subsubsection}{\large\bfseries}
\titleformat*{\paragraph}{\large\bfseries}
\titleformat*{\subparagraph}{\large\bfseries}

\usepackage[a4paper, total={6in, 8in}]{geometry}
\begin{document}
\title{User Manual for the SU2 EQUiPS Module: Enabling Quantification of Uncertainty in Physics Simulations}
\author{Jayant Mukhopadhaya}
\address{Department of Aeronautics \& Astronautics, Stanford University, Stanford, California-94305, USA}

\maketitle

\chapter{Introduction}
This document serves as the manual for using the EQUiPS (Enabling Quantification of Uncertainty in Physics Simulations) module in SU2. The EQUiPS module uses the Eigenspace Perturbation methodology \cite{eigper} to provide interval bounds on Quantities of Interest (QoIs) that capture epistemic uncertainties arising from assumptions made in RANS turbulence models. This has been implemented and tested in SU2 \cite{mishra2019uncertainty} for a variety of benchmark turbulence cases as well as flows of aerodynamic interest. 

One of the key features of the EQUiPS module that we strove to achieve in its implementation is \textit{versatility}: ensuring that anyone, regardless of their background in turbulence modeling, can utilize this module. The module can be used without explicit knowledge of the physics behind the methodology but some basic details are needed to understand the outputs of the simulations, and how to create the interval bounds on quantities of interest. 

The methodology requires 5 perturbed simulations, in addition to a baseline unperturbed simulation, to characterize the epistemic uncertainties due to turbulence modeling. This baseline simulation refers to the regular turbulent RANS simulation that would be performed in SU2. The perturbed simulations are performed sequentially by a python script, the instructions for which are detailed in chapter \ref{python}. Each perturbed simulation results in a different realization of the flow field, and by extension, a different realization of the QoIs. The interval bounds are formed by the maximum and minimum values the QoIs resulting from these 6 simulations. It is important to note that the UQ functionality is only available for the $k-\omega$ SST turbulence model, at present. 

This manual starts by explaining the theory underlying the Eigenspace Perturbation Framework that is implemented in the EQUiPS module. Then, it walks the reader through the process of installing SU2 in chapter \ref{installation}. This is followed by instructions on using the EQUiPS module by either running all the perturbed simulations sequentially in chapter \ref{python}, or running them individually in chapter \ref{options}. These instructions are made concrete with examples included in chapter \ref{ex1}. Finally, we highlight the use of this module in published literature in chapter \ref{usage}.

\chapter{Introduction to the Eigenspace Perturbation Framework} \label{theory}

In this chapter, we provide the mathematical and computational background for the Eigenspace perturbation methodology. We introduce each sequentially, starting from the eigen-decomposition of the modeled Reynolds stresses, the introduction of the perturbations into the eigenvalues and eigenvectors to account for structural uncertainty, and deriving the uncertainty estimates from the perturbed simulations.

\section*{ Representation of the uncertainty: Eigenspace expression of Reynolds stresses}
The Reynolds stress tensor, $R_{ij}=\langle u_iu_j \rangle$, is a primary Quantity of Interest for turbulence modeling. This can be decomposed into facets that determine the shape, the orientation and the amplitude of the Reynolds stress ellipsoid. To this end, the Reynolds stress tensor can be decomposed into the anisotropic and deviatoric components as
\begin{equation}
R_{ij}=2k(b_{ij}+\frac{\delta_{ij}}{3}).
\end{equation}
Here, $k(=\frac{R_{ii}}{2})$ is the turbulent kinetic energy and $b_{ij}(=\frac{R_{ij}}{2k}-\frac{\delta_{ij}}{3})$ is the Reynolds stress anisotropy tensor. The Reynolds stress anisotropy tensor can be expressed as
\begin{equation}
b_{in}v_{nl}=v_{in}\Lambda_{nl},
\end{equation}
where $v_{nl}$ is the matrix of orthonormal eigenvectors and $\Lambda_{nl}$ is the traceless diagonal matrix of eigenvalues $\lambda_{k}$. Multiplication by $v_{jl}$ yields $b_{ij}=v_{in}\Lambda_{nl}v_{jl}$. This is substituted into Equation (1) to yield
\begin{equation}
R_{ij}=2k(v_{in}\Lambda_{nl}v_{jl}+\frac{\delta_{ij}}{3}).
\end{equation}
The tensors $v$ and $\Lambda$ are ordered such that $\lambda_{1}\geq\lambda_{2}\geq\lambda_{3}$. In this representation, the shape, the orientation and the amplitude of the Reynolds stress ellipsoid are directly represented by the turbulence anisotropy eigenvalues $\lambda_l$,  eigenvectors $v_{ij}$ and the turbulent kinetic energy $k$, respectively.

The limitations of classical turbulence models can be re-expressed using this decomposition. For instance, one of the key ramifications of the eddy-viscosity hypothesis is that it obligates the modeled Reynolds stress to share its eigen-directions with the mean rate of strain tensor. Consequently, the eigenvectors of the modeled Reynolds stresses are co-incident with those of the mean rate of strain. While this is true in simple shear flows, it is limited in complex engineering flows. Similarly, assumptions made in the gradient diffusion hypothesis lead to imperfect representation of the amplitude of the Reynolds stress ellipsoid and the form of the eddy-viscosity hypothesis leads to unsatisfactory expression for the Reynolds stress anisotropy eigenvalues. 

\section*{ Application of the perturbations: Eigenvalue \& Eigenvector perturbations}
To account for the errors due to closure assumptions, this eigenspace representation of the Reynolds stress tensor is perturbed. These perturbations are injected directly into the modeled Reynolds stress during the CFD solution iterations. This perturbed form is expressed as:
\begin{equation}
R_{ij}^*=2k^* (\frac{\delta_{ij}}{3}+v^*_{in}\Lambda^*_{nl}v^*_{lj})
\end{equation}
where $^*$ represents the perturbed quantities. The perturbations to the eigenvalues, $\Lambda$, correspond to varying the componentiality of the flow (or the shape of the Reynolds stress ellipsoid). Similarly, the perturbations to the eigenvectors and the turbulent kinetic energy vary the orientation and amplitude of the Reynolds stress ellipsoid. These perturbations are sequentially applied to the modeled Reynolds stress tensor.

\begin{figure}
\center
\includegraphics[width=\textwidth]{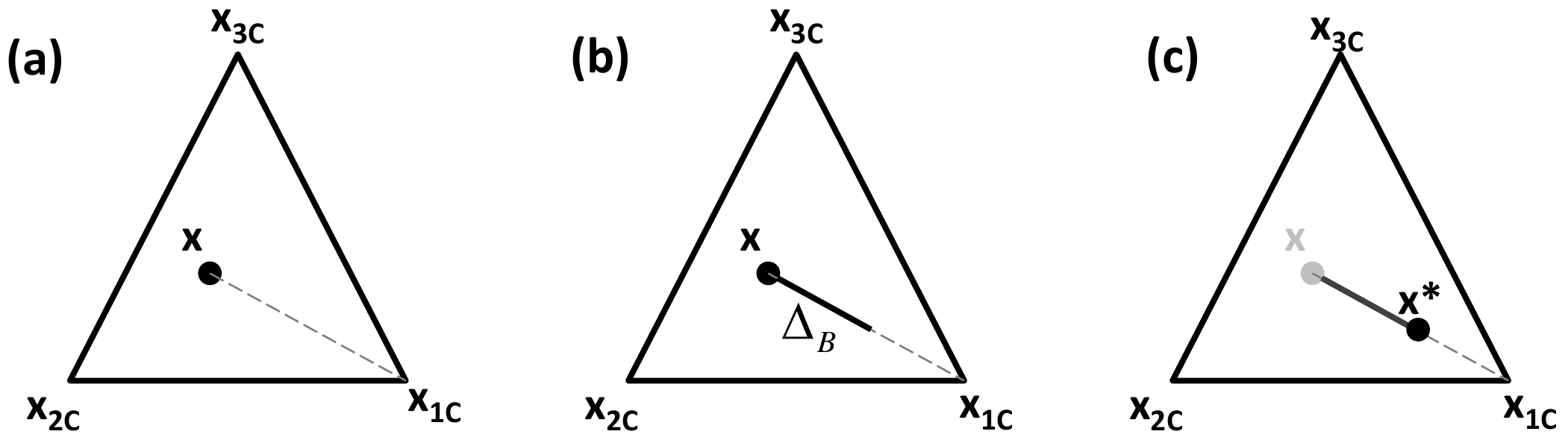}
\caption{Schematic outline of the eigenvalue perturbations on the barycentric triangle, starting from an arbitrary state. \label{fig:eigenvalueillustration}}
\end{figure}

\textit{Eigenvalue perturbation}: The eigenvalue perturbation can be represented on the barycentric map. In this representation, all realizable states of the Reynolds stress tensor lie on or inside the barycentric triangle. The vertices of this triangle, labeled $\mathbf{x_{1C}}$, $\mathbf{x_{2C}}$ and $\mathbf{x_{3C}}$ in Fig. \ref{fig:eigenvalueillustration}, represent the one, two and three component limiting states of the turbulent flow field. A linear map between the co-ordinates on this triangle $\mathbf{x}$ and the Reynolds stress anisotropy eigenvalues $\lambda_i$ is defined by
\begin{equation}
\mathbf{x}=\mathbf{x_{1C}}(\lambda_1 - \lambda_2) + \mathbf{x_{2C}}(2\lambda_2 - 2\lambda_3) +\mathbf{x_{3C}}(3\lambda_3- 1). 
\end{equation}
This linear transformation can be expressed as $\mathbf{x}=\mathbf{B}\lambda$. In physical terms, this invertible, one-to-one mapping expresses any realizable state of the Reynolds stress eigenvalues as a convex combination of the three limiting states of turbulence. 

The projection of the eigenvalue perturbation in the barycentric map has both a direction and a magnitude, as is exhibited in Fig. \ref{fig:eigenvalueillustration}. In this application, the perturbations are aligned towards the vertices of the barycentric triangle (or the limiting states of turbulence), as shown in in Fig. \ref{fig:eigenvalueillustration}. The magnitude of the eigenvalue perturbation in the barycentric triangle is represented by $\Delta_B \in [0,1]$, schematically illustrated in Fig. \ref{fig:eigenvalueillustration} (b). The perturbed barycentric coordinates $\mathbf{x^*}$ are given by
$\mathbf{x^*}=\mathbf{x}+\Delta_B(\mathbf{x^{t}}-\mathbf{x})$, where $\mathbf{x^{t}}$ denotes the target vertex (representing one of the one-, two-, or three-component limiting
states) and $\mathbf{x}$ is the unperturbed model prediction. Thus, $\Delta_B=0$ would leave the state unperturbed and $\Delta_B=1$ would perturb any arbitrary state to the vertices of the barycentric triangle. In Fig. \ref{fig:eigenvalueillustration}, this eigenvalue perturbation methodology is illustrated, starting from an arbitrary Reynolds stress componentiality. For this illustration, the direction of the perturbation $\mathbf{x^{t}}$ is chosen toward $\mathbf{x_{1C}}$ and the magnitude of perturbation $\Delta_B $ is chosen as $0.5$. The initial $\mathbf{x}$ and perturbed $\mathbf{x^*}$  states are exhibited in the figure, along with the transition. 

\textit{Eigenvector perturbations}: The eigenvector perturbations vary the alignment of the Reynolds stress ellipsoid. These are guided by the turbulence production mechanism, $\mathcal{P}=-R_{ij}\frac{\partial U_i}{\partial x_j}$. The eigenvector perturbations seek to modulate turbulence production by varying the Frobenius inner product $\langle A,R \rangle =tr(AR)$, where $A$ is the mean velocity gradient and $R$ is the Reynolds stress tensor. For the purposes of bounding all permissible dynamics, we seek the extremal values of this inner product. In the coordinate system defined by the eigenvectors of the rate of strain tensor, the critical alignments of the Reynolds stress eigenvectors are given by $v_{max}=\begin{bmatrix}
  1 & 0 & 0 \\
  0 & 1 & 0 \\
  0 & 0 & 1
 \end{bmatrix}$  and $v_{min}=\begin{bmatrix}
  0 & 0 & 1 \\
  0 & 1 & 0 \\
  1 & 0 & 0
 \end{bmatrix}$. The range of this inner product is $[\lambda_1\gamma_3+\lambda_2\gamma_2+\lambda_3\gamma_1, \lambda_1\gamma_1+\lambda_2\gamma_2+\lambda_3\gamma_3]$, where $\gamma_1 \geq \gamma_2 \geq \gamma_3$ are the eigenvalues of the symmetric component of $A$. 
 
 \begin{figure}
\center
\includegraphics[width=\textwidth]{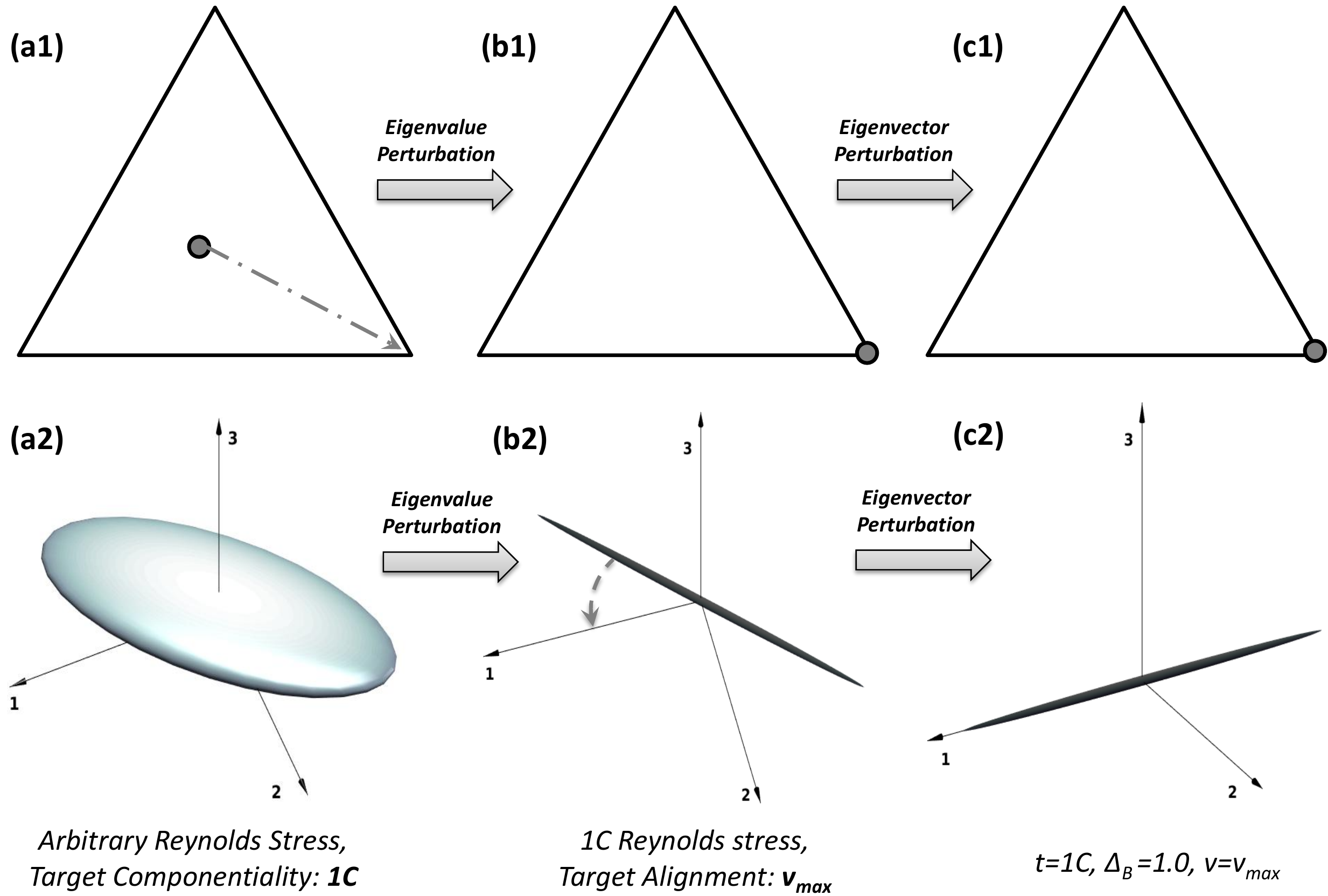}	
\caption{Schematic outline of Eigenspace perturbations from an arbitrary state of the Reynolds stress. \label{fig:eigenspaceillustration}}
\end{figure}

In physical terms, the eigenvalue perturbation changes the shape of the Reynolds stress ellipsoid and the eigenvector perturbation changes its relative alignment with the principal axes of the mean rate of strain tensor. To illustrate this eigenspace perturbation framework, we outline a representative case schematically in Fig. \ref{fig:eigenspaceillustration}. In the upper row, $(a_1,b_1,c_1)$, we represent the Reynolds stress tensor at a specific physical location in barycentric coordinates and in the lower row, $(a_2,b_2,c_2)$, we visualize the Reynolds stress ellipsoid in a coordinate system defined by the mean rate of strain eigenvectors. These are arranged so that $\lambda_1 \geq \lambda_2 \geq \lambda_3,$. Thus, the 1-axis is the stretching eigendirection and the 3-axis is the compressive eigendirection of the mean velocity gradient.
 
Initially, the Reynolds stress predicted by an arbitrary model is exhibited in the first column, Fig. \ref{fig:eigenspaceillustration}, $a1$ and $a2$. The eigenvalue perturbation methodology seeks to sample from the extremal states of the possible Reynolds stress componentiality. Thus, we may, for instance, translate the Reynolds stress from this state to the $1C$ state, exhibited in the transition from Fig. \ref{fig:eigenspaceillustration}, $a1$ to $b1$. This translation changes the shape of the Reynolds stress ellipsoid from a a tri-axial ellipsoid to a prolate ellipsoid, exhibited in the transition from Fig. \ref{fig:eigenspaceillustration}, $a2$ to $b2$.
 
Thence, the eigenvector perturbation varies the alignment of this ellipsoid. Thus, we may, for instance, rotate the Reynolds stress ellipsoid so that its semi-major axis is aligned with the stretching eigendirection of the mean rate of strain tensor, exhibited in the transition from Fig. \ref{fig:eigenspaceillustration}, $b2$ to $c2$. This particular alignment would enable us to analyze impact of the maximum permissible production on turbulence evolution. In conjunction, these two perturbation approaches enable us to maximize the information we may get from single-point statistics to quantify uncertainty estimates.

\begin{figure}
\center
\includegraphics[width=\textwidth]{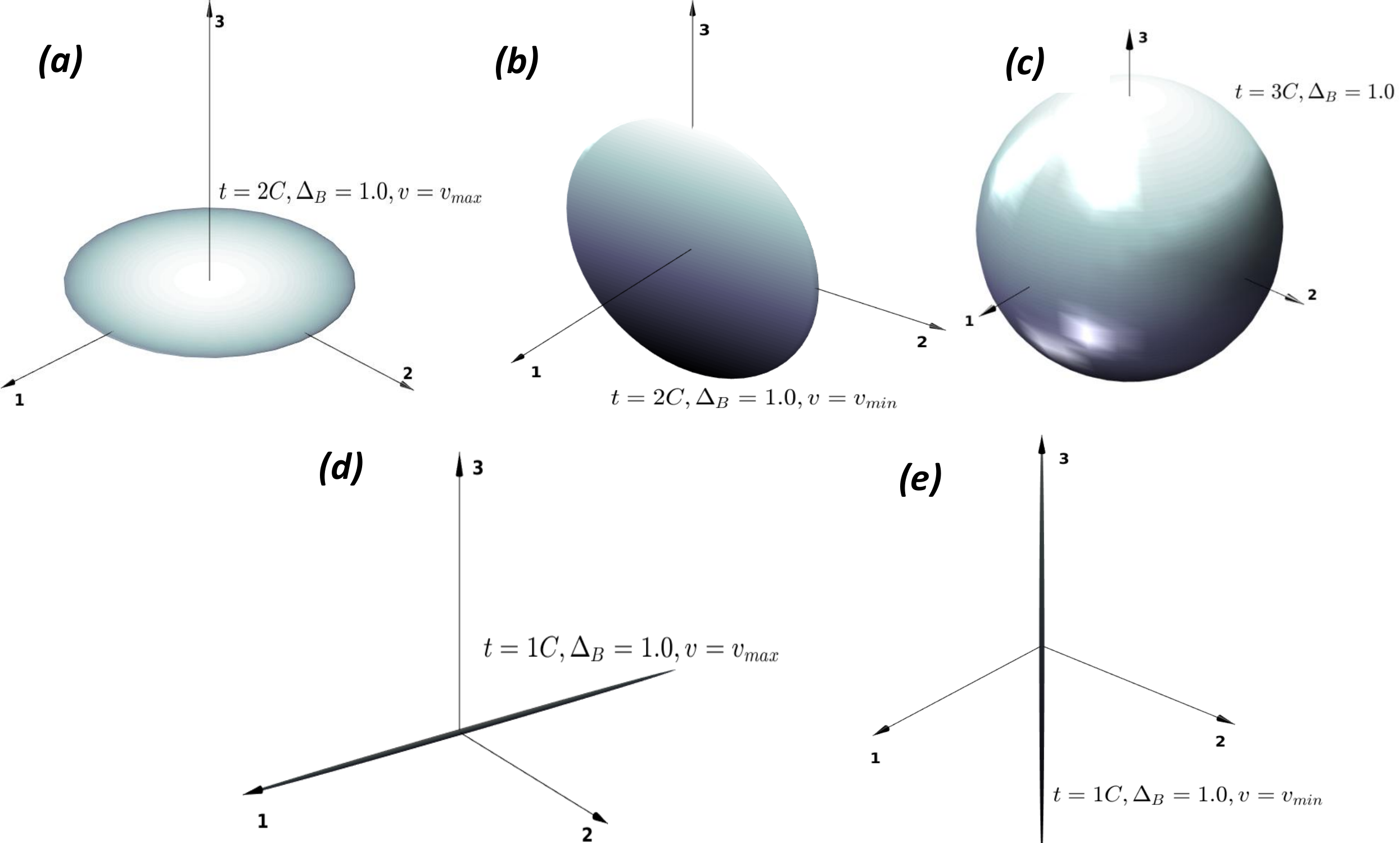}
\caption{Schematic visualization of the extremal states, as Reynolds stress ellipsoids, in the eigenspace perturbation methodology.\label{fig:states}}
\end{figure}

This eigenspace perturbation framework gives us $5$ distinct extremal states of the Reynolds stress tensor, these are schematically displayed in Fig. \ref{fig:states}. These correspond to 3 extremal states of the componentiality $(1C, 2C, 3C)$ and 2 extremal alignments of the Reynolds stress eigenvectors, $(v_{min}, v_{max})$.  For the $3C$ limiting state, the Reynolds stress ellipsoid is spherical. Due to rotational symmetry, all alignments of this spherical Reynolds stress ellipsoid are identical and eigenvector perturbations are superfluous.  

\section*{Determining the uncertainty: Uncertainty estimates}
\begin{figure}
\center
\includegraphics[angle=90,origin=c,width=0.48\textwidth]{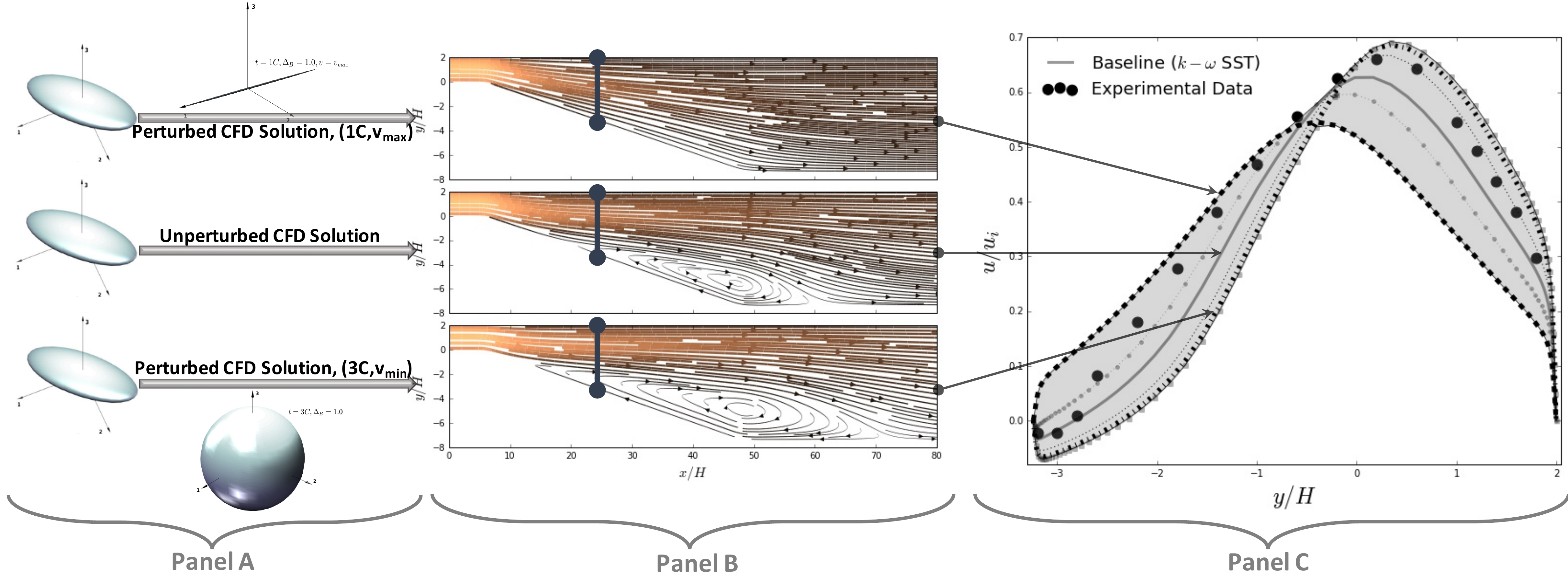}
\caption{Schematic outlining stages in computation of uncertainty estimates. \textit{Panel A}:RANS simulations with perturbations; \textit{Panel B}:perturbed realizations of turbulent flow fields; \textit{Panel C}:Compositions of uncertainty estimates from union of perturbed QoI profiles.\label{fig:boundsillustration}}
\end{figure}

In this subsection, we outline how the uncertainty estimates are engendered from the set of perturbed CFD simulations. This process is schematically exhibited in Fig. \ref{fig:boundsillustration}. The illustrative flow used is the canonical case of separated turbulent flow in a planar diffuser. The conditions and the experimental data are from the experimental study of \cite{buice}. 

The central panel of Fig. \ref{fig:boundsillustration} outlines the unperturbed, baseline CFD solution. Using the $k-\omega$ SST model, this leads to a unique flow field realization in the flow domain. To illustrate the composition of the uncertainty bounds, we choose a specific location in the domain, specifically at $x/H=24$ which is marked in the figures. This unique flow field realization from the SST model leads to a singleton profile for the mean velocity, $u_i/u$, shown in panel C with the solid gray line. 

The upper and lower panels of the figure outline perturbed solutions. While there are 5 perturbed states as discussed in the last subsection, we exhibit only 2 of these in the illustration. Each of these perturbed solutions leads to a different realization of the flow field, as is illustrated in panel B. These flow realizations differ in essential aspects. For instance, the perturbation to the state $(1C, v_{max})$ maximizes the turbulence production mechanism and thus, suppresses flow separation. The perturbation to the state $(3C, v_{min})$ minimizes the turbulence production mechanism and thus, strengthens flow separation. This is evidenced in the variation of the separation zones in panel B. Each of these perturbations leads to a different flow field and consequently, the velocity profiles from these flow fields are different as well. The velocity profiles at $x/H=24$ from the  $(1C, v_{max})$ and $(3C, v_{min})$ are shown in panel C with the dashed and dot-dashed lines respectively. (panel C also shows the profiles from the $(1C, v_{min})$, $(2C, v_{max})$ and $(2C, v_{min})$ perturbations using the dotted line, dotted line with circles and dotted lines with squares) The uncertainty estimates on the profiles of a quantity of interest (QoI) at a location are engendered by the union of all the
states lying in the profiles from this set of perturbed RANS simulations. This is illustrated by the gray shaded zone in Fig. \ref{fig:boundsillustration}  panel C. 

At this juncture, it may be useful to outline what the uncertainty estimates are, and more importantly, what they are not. These uncertainty estimates do not represent confidence or prediction intervals at any significance level. To generate such confidence or prediction intervals, one may require data from high-fidelity realizations of the flow (Direct Numerical Simulation (DNS), Large Eddy Simulation (LES) or experimental studies) along with assumptions regarding the distributions of the explanatory features. In a stereotypical design investigation, one has access to little or no high-fidelity data. Additionally, any assumptions about the distribution would require knowledge of for instance the history of the turbulent flow field, which is not feasible for most engineering applications. The uncertainty estimates outlined in this investigation are \textit{data-free} and rely on a purely physics-based framework. In this context, the uncertainty estimates represent estimated \textit{ranges} for the values of a Quantity of Interest contingent upon the uncertainties and discrepancies arising due to the model-form of eddy-viscosity based RANS closures. 

\chapter{Downloading and Installing the EQUiPS Module} \label{installation}

The EQUiPS module is implemented in \href{https://su2code.github.io/}{SU2}, an open-source multi-physics simulation software. The software binaries can be downloaded from the \href{https://su2code.github.io/download.html}{website}. For best performance, consider building SU2 from the source code. The repository for SU2 is hosted on GitHub. The best way to download the source code would be to clone the repository with the command: 

\noindent \texttt{git clone https://github.com/su2code/SU2.git}

If you have already cloned the SU2 repository in the past, it is a good idea to update the current version with the latest changes in the remote repository using

\noindent\texttt{git pull}

Detailed instructions to build SU2 from the source code are available \href{https://su2code.github.io/docs_v7/Build-SU2-Linux-MacOS/}{here}. 

\chapter{Running the Python Script} \label{python}

As mentioned in the Introduction, the python script sequentially performs the 5 perturbed simulations that are required to inform the interval bounds. For smooth operation, it is best to have performed a baseline unperturbed simulation with SU2 and have achieved sufficient convergence. This ensures that the mesh file, and the input configuration file are well posed, and, if run through the Python script, can provide converged perturbed solutions. 

The Python script takes an input configuration file that identical to the one used to run the baseline CFD simulation in SU2. The options for the python script, and their uses are:

\begin{itemize}
\item \texttt{-f}: Specifies the name of the configuration file to be used
\item \texttt{-n}: Sets the number of processors being used to run the simulations. A parallel build is required to use this option
\item \texttt{-u}: Sets the under-relaxation factor used in performing perturbation. This option need not be changed unless the perturbation simulations are unstable. $u \in [0,1]$ and it's default value is $0.1$. This should not be set to $< 0.05$ as the perturbations may not be completed by convergence. 
\item \texttt{-b}: Sets the magnitude of perturbation. This option should not be touched without having read the references on the Eigenspace Perturbation methodology \cite{eigper,mishra2019uncertainty}. $b \in [0,1]$ and it's default value is $1.0$. The default value corresponds to a full perturbation and is required to correctly characterize the epistemic uncertainties
\end{itemize}

The most common use of this script would be: 

\texttt{compute\_uncertainty.py -f turb\_naca0012.cfg -n 8}

This will run the 5 perturbed simulations for the case defined in the \texttt{turb\_naca0012.cfg} configuration file on 8 processors. It creates a new directory for each new simulation, and outputs the results in the respective directories. The directories are named: \texttt{1c, 2c, 3c, p1c1} and \texttt{p1c2}. Each flow solution is an instantiation of the flow field that must be post-processed to extract the necessary model form uncertainty information. 

It is important to note that this UQ functionality is only available with the SST turbulence model. 

\chapter{Running Individual Perturbations} \label{options}

The python script is an easy interface to run the perturbed simulations sequentially. In case there is a need to perform the simulations separately (for example to run them in parallel, or on different machines), individual perturbations can be performed by setting options within the configuration file. The list of the different options available is given below. 

\begin{itemize}
\item \texttt{USING\_UQ}: Boolean that ensures EQUiPS module is used
\item \texttt{UQ\_COMPONENT}: Number that specifies the eigenvalue perturbation to be performed
\item \texttt{UQ\_PERMUTE}: Boolean that indicates whether eigenvector permutation needs to be performed
\item \texttt{UQ\_URLX}: Sets the under-relaxation factor used in performing perturbation. This option need not be changed unless the perturbation simulations are unstable. $u \in [0,1]$ and it's default value is $0.1$. This should not be set to $< 0.05$ as the perturbations may not be completed by convergence. 
\item \texttt{UQ\_DELTA\_B}: Sets the magnitude of perturbation. This option should not be touched without having read the references on the Eigenspace Perturbation methodology \cite{eigper,mishra2019uncertainty}. $\Delta_b \in [0,1]$ and it's default value is $1.0$. The default value corresponds to a full perturbation and is required to correctly characterize the epistemic uncertainties
\end{itemize}

An example of how the configuration options would look, is shown below:

\begin{lstlisting}[frame=single]
% ------------------ UNCERTAINTY QUANTIFICATION DEFINITION ------------------%
%
% Using uncertainty quantification module (YES, NO). Only available with SST
USING_UQ= YES
%
% Eigenvalue perturbation definition (1, 2, or 3)
UQ_COMPONENT= 1
%
% Permuting eigenvectors (YES, NO)
UQ_PERMUTE= NO
%
% Under-relaxation factor (float [0,1], default = 0.1)
UQ_URLX= 0.1
%
% Perturbation magnitude (float [0,1], default= 1.0)
UQ_DELTA_B= 1.0
\end{lstlisting}

Even though each perturbed simulation can be performed individually, all 5 perturbed simulations, in addition to the baseline unperturbed simulation are required to characterize the interval bounds. A combination of \texttt{UQ\_COMPONENT} and \texttt{UQ\_PERMUTE} options are required to perform the 5 different perturbations. These combinations are enumerated in Table \ref{perturbationTable}. These simulations can be run independently from each other which allows for parallelization of the simulations. This also allows for the tuning of the convergence parameters for the different simulations.

\begin{table}
\caption{\label{perturbationTable}Combination of options required to perform perturbed simulations}
\begin{tabular}{ |c|c|c| } 
 \hline
 Perturbation & \texttt{UQ\_COMPONENT} & \texttt{UQ\_PERMUTE} \\
 \hline
 \texttt{1c} & 1 & \texttt{NO} \\
 \texttt{2c} & 2 & \texttt{NO} \\
 \texttt{3c} & 3 & \texttt{NO} \\
 \texttt{p1c1} & 1 & \texttt{YES} \\
 \texttt{p1c2} & 2 & \texttt{YES} \\
 \hline
\end{tabular}
\end{table}

\chapter{NACA0012 Example} \label{ex1}

To illustrate the capabilities of the EQUiPS module, some simple test cases are explored. The first test case concerns flow over a NACA0012 airfoil at a range of angles attack from $0^\circ$ to $20^\circ$. This is a simple 2D geometry that stalls, and exhibits separated flow, at high angles of attack. It is a ubiquitous geometry that has significant amounts of experimental data available that allows for the comparison of lower fidelity RANS CFD simulations, to the higher fidelity wind tunnel tests that have been conducted.

This tutorial is also available through the \href{https://su2code.github.io/tutorials/UQ_NACA0012/}{SU2 website} and the relevant configuration and mesh files are available from the \href{https://github.com/su2code/Tutorials/tree/master/compressible_flow/UQ_NACA0012}{GitHub repository}.

\section*{Problem setup}
This problem will solve the flow past the airfoil with the conditions shown in Table \ref{naca0012_test_cond}.

\begin{table}
\centering
    \captionsetup{justification=centering}
    \caption{Simulation conditions for the NASA CRM.} 
    \begin{tabular}{|c|c|}
        \hline
        Mach Number & $0.15$ \\ \hline
        Reynolds Number & $6\times10^6$ \\ \hline
        Reference chord length & $1.0$ m \\ \hline
        Freestream Temperature & $300.0~\text{K}$ \\ \hline
        $\alpha$ & $-2^\circ \leq \alpha \leq 12^\circ$ \\  
        \hline
    \end{tabular}
    \label{naca0012_test_cond}
\end{table}


Although this particular case simulates flow at $15^\circ$, the same simulation can be run at varying angles of attack. The results section also presents analyses from performing the simulations at a range of angles of attack which allows the exploration of the various flow regimes that occur. At low angles of attack, the flow stays attached and RANS simulations are quite accurate in predicting the flow. At higher angles of attack, the onset of stall causes flow separation which leads to inaccuracies in flow predictions.

\section*{Mesh Description}
\begin{figure}
\center
\includegraphics[width=0.75\textwidth]{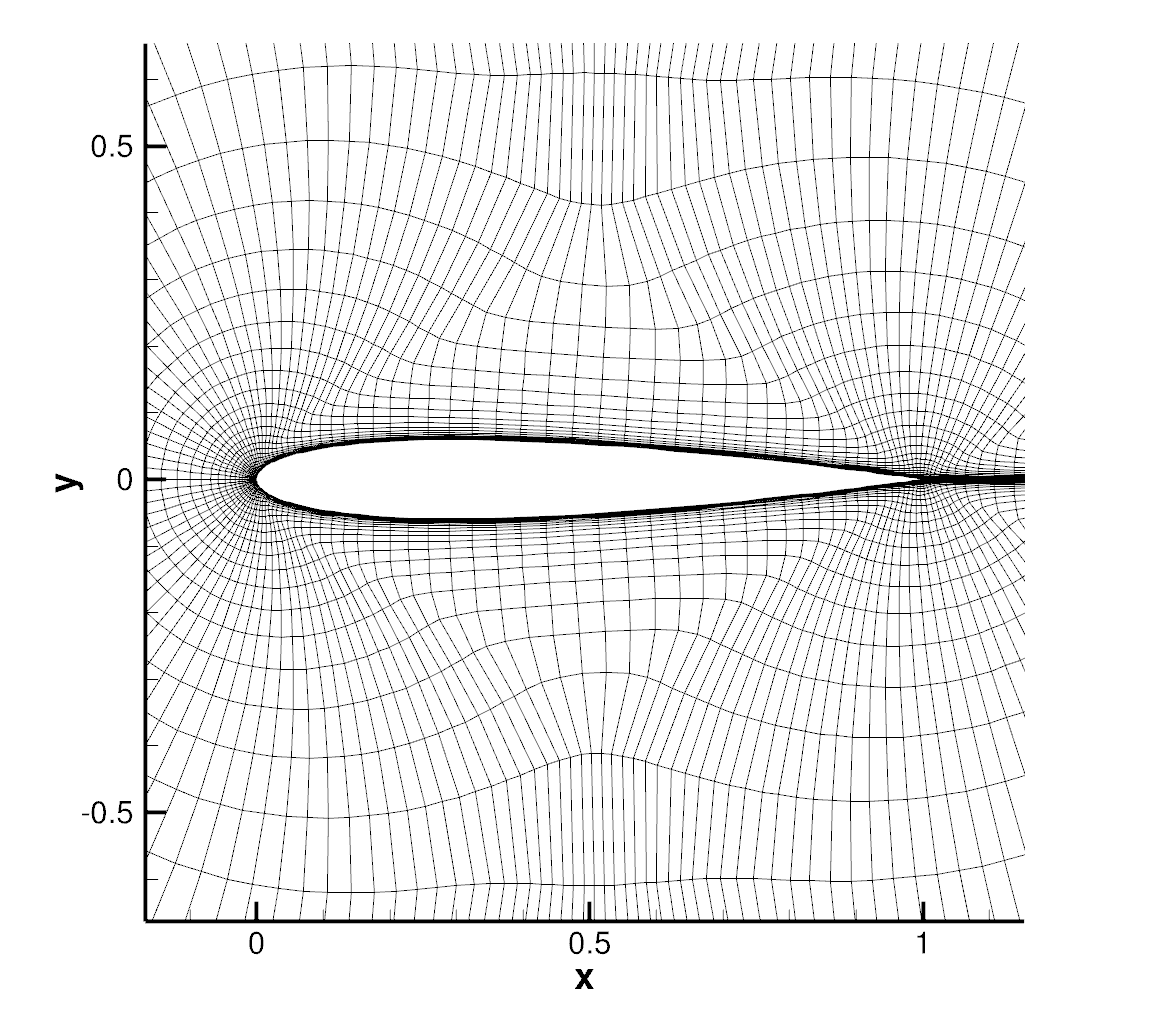}
\caption{Magnified view of the NACA0012 mesh near the wall \label{fig:n0012_mesh}}
\vspace{-2mm}
\end{figure}

The mesh is a structured C-grid. The farfield boundary extends 500c away from the airfoil surface. A magnified view of the mesh near the wall can be seen in Fig. \ref{fig:n0012_mesh}.

\section*{Running the Module}
The module is built to be versatile, such that it can be used by experts and non-experts alike. A simple Python script abstracts away the details of the perturbations (componentality, eigenvector permutations) and sequentially performs the perturbed simulations. The script requires a mesh and configuration file that are identical to ones that are needed to run a baseline RANS CFD simulation. For smooth operation, it is best to have performed the baseline simulation with SU2 and have achieved sufficient convergence. This ensures that the configuration file and mesh are well posed, and, if run through the Python script, can provide converged, perturbed simulations. Details in the next section on Configuration File Options are not required to run the Python script. Unless there is a need to perform the perturbations individually, you can move to the Running SU2 section.

\section*{Configuration File Options}
If there is a need to perform the perturbations individually (for example to run them in parallel, or on different machines), configuration options need to be set to specify the kind of perturbation to perform.

\newpage
\begin{lstlisting}[frame=single]
% ------------------- UNCERTAINTY QUANTIFICATION DEFINITION -------------------%
% Using uncertainty quantification module (YES, NO). Only available with SST
USING_UQ= YES
%
% Eigenvalue perturbation definition (1, 2, or 3)
UQ_COMPONENT= 1
%
% Permuting eigenvectors (YES, NO)
UQ_PERMUTE= NO
%
% Under-relaxation factor (float [0,1], default = 0.1)
UQ_URLX= 0.1
%
% Perturbation magnitude (float [0,1], default= 1.0)
UQ_DELTA_B= 1.0
\end{lstlisting}

\section*{Results}
\begin{figure}
\center
\includegraphics[trim=80 180 115 200, clip, width=0.6\textwidth]{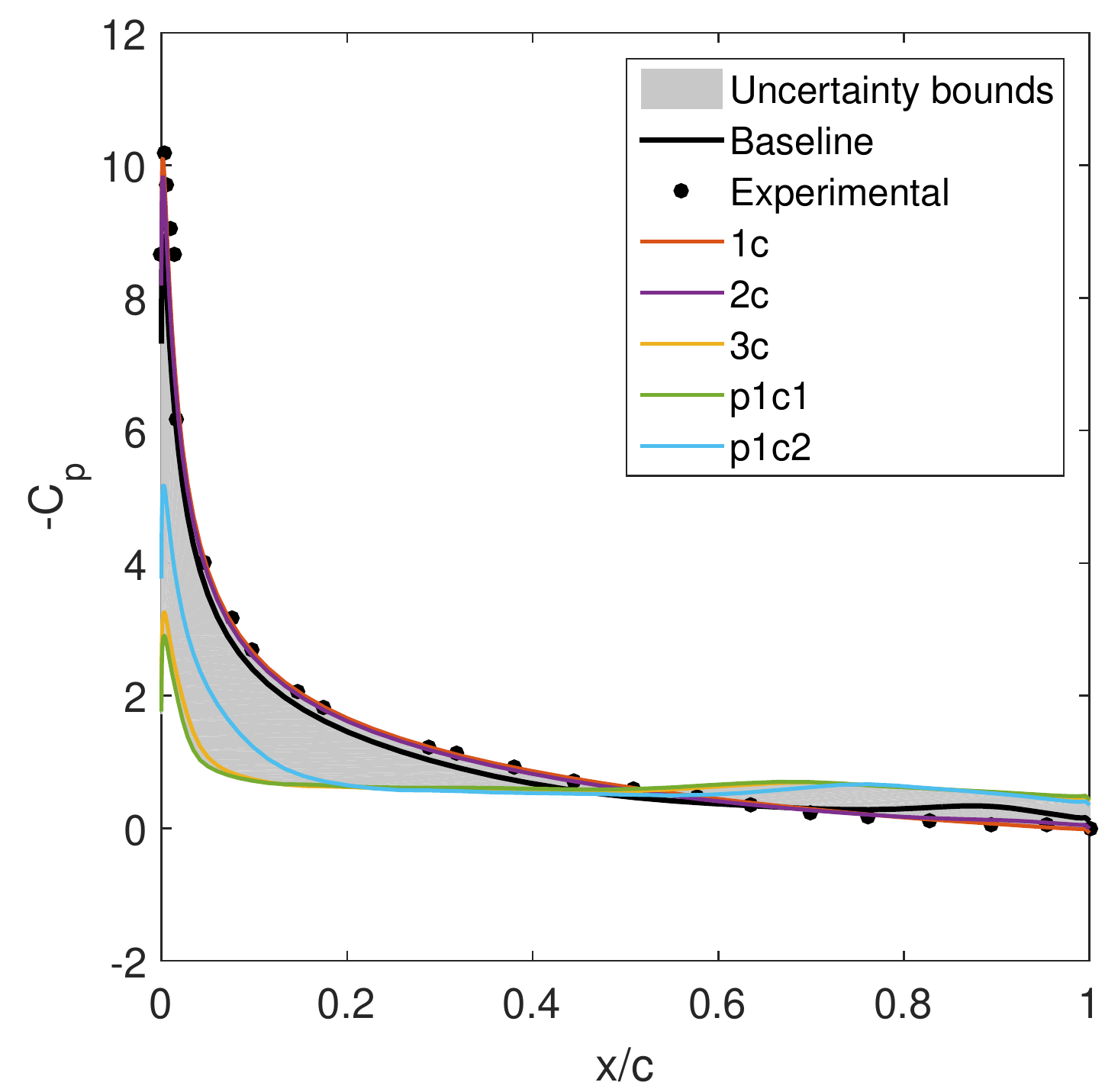}
\includegraphics[trim=80 180 115 200, clip, width=0.6\textwidth]{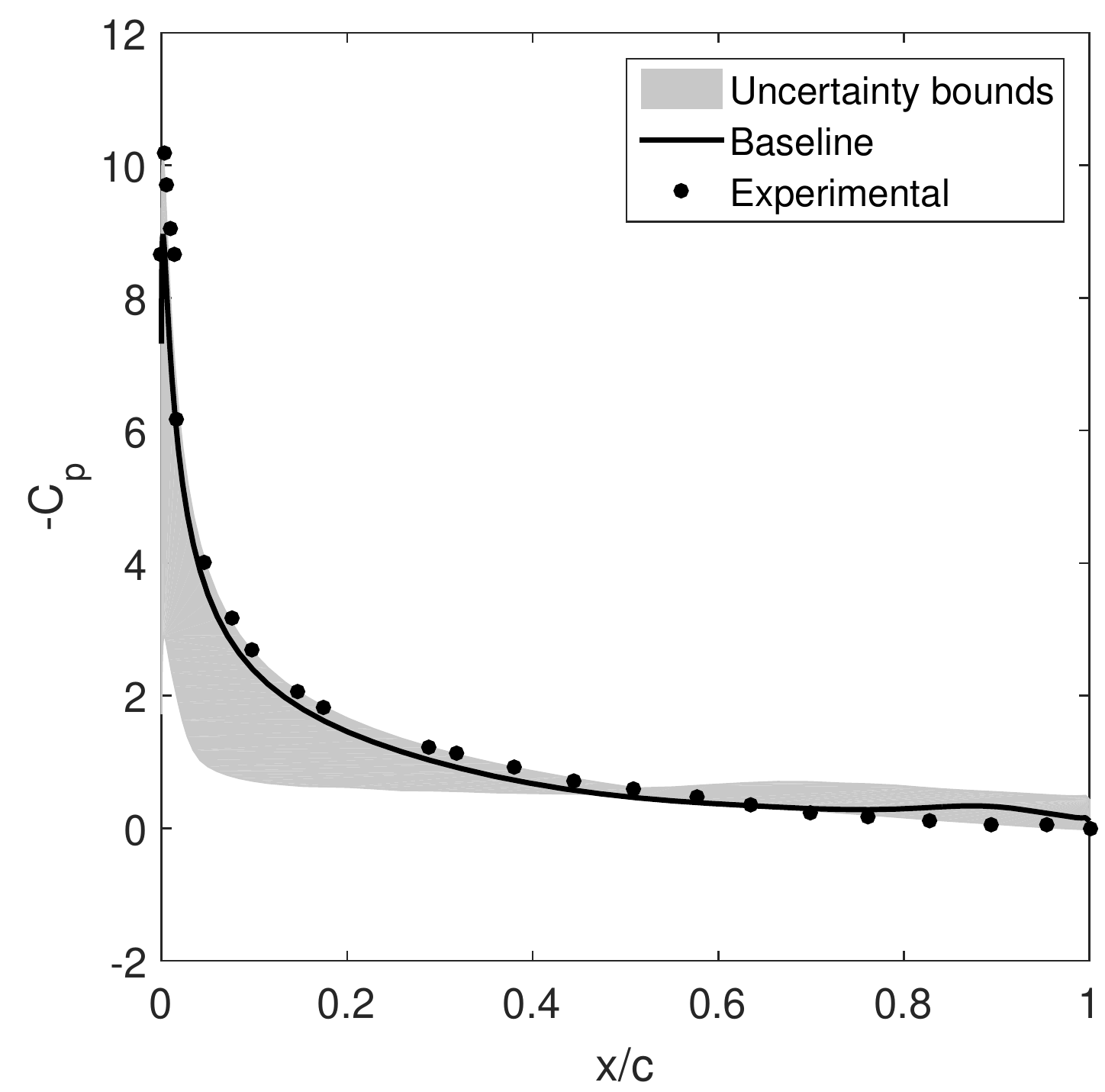}
\caption{$C_P$ distribution along upper surface for the NACA0012 airfoil at 15deg AOA (a) with individual perturbations included, (b) with only the resulting interval bounds.\label{fig:naca0012_cp}}
\end{figure}

In order to obtain the interval bounds of a QOI, all 6 instantiations of the flow solution (1 baseline and 5 perturbed) must be analyzed. To illustrate how the bounds are formed, we use the example of the $C_P$ distribution along the upper surface of the airfoil. In Fig. \ref{fig:naca0012_cp}(a) the $C_P$ distributions of each perturbed simulation is plotted along with the baseline simulation, experimental data, and the uncertainty bounds. In Fig. \ref{fig:naca0012_cp}(b), only the individual perturbation data is hidden. The uncertainty bounds are formed by a union of all the states the QOI predicted by the module. It is interesting to see the bounds are larger in areas with correspondingly large discrepancy between the baseline simulation, and the experimental data.

As we can see in Fig. \ref{fig:naca0012_cp}(b), the predictions of the RANS model are not in perfect agreement with the experimental data. This lack of agreement is even more severe near $x/c=0$. However, the uncertainty estimates from the EQUiPS module account for this discrepancy and the experimental data lies in the uncertainty estimates. Analysing further from Fig. \ref{fig:naca0012_cp}(a), we observe that the experimental data are in agreement with the $1C$ perturbations. These represent limiting states of the Reynolds stress anisotropy and eddy-viscosity based models are not able to predict such extreme states of anisotropy as their predictions are restricted to the plane strain line of the barycentric triangle. 

\begin{figure}
\center
\includegraphics[trim=80 180 115 200, clip, width=0.6\textwidth]{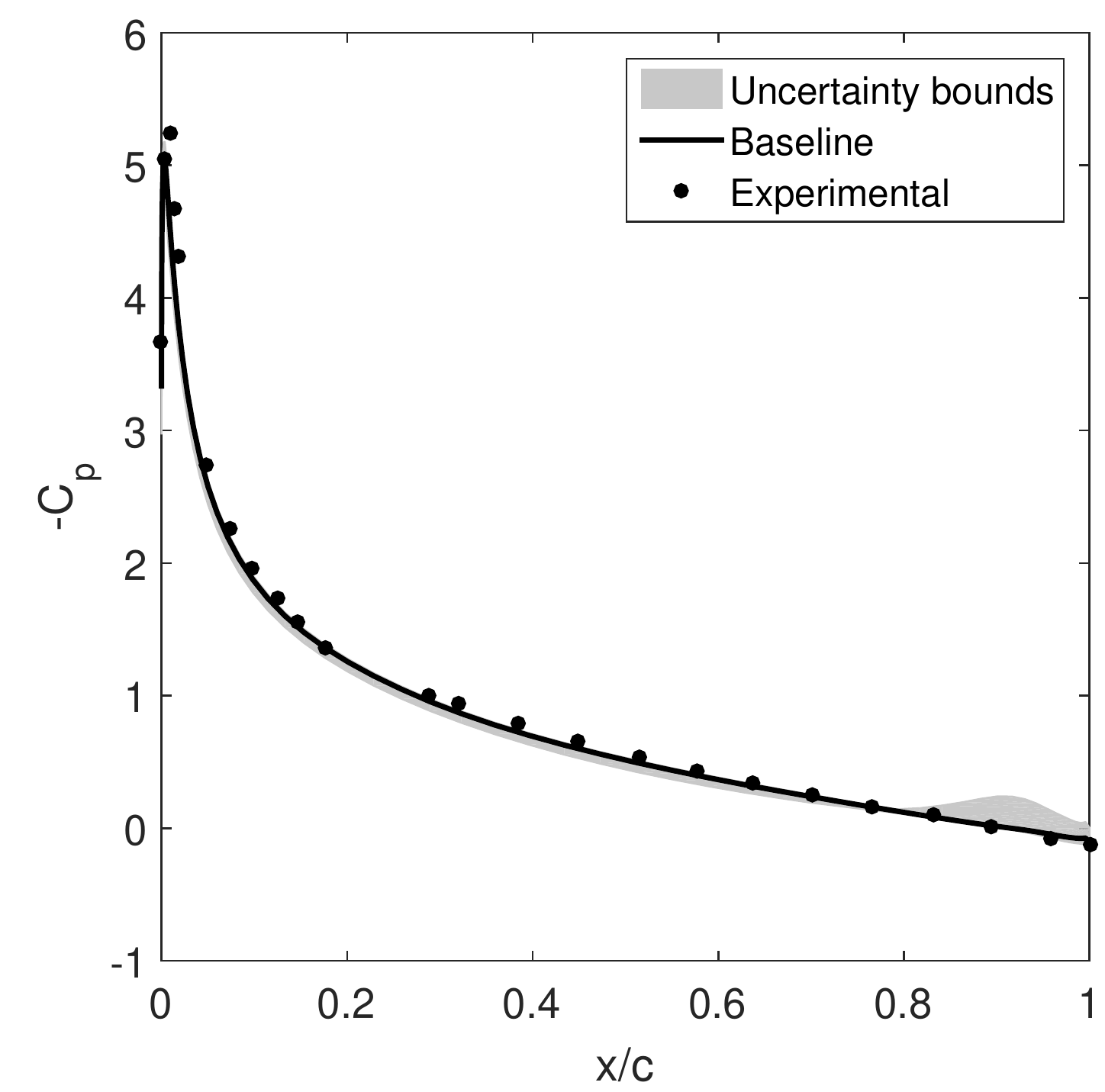}
\caption{$C_P$ distribution along upper surface for the NACA0012 airfoil at 10deg AOA with predicted interval bounds \label{fig:naca0012_cp_10}}
\end{figure}

At an angle of attack of 10deg, the baseline RANS model is able to accurately predict the $C_P$ distribution. If the UQ module is run at this angle, it is seen that the uncertainty bounds are much smaller. This case can be run simply using the steps as above, only changing the AOA option for the files. This is illustrated in Fig. \ref{fig:naca0012_cp_10}.

\begin{figure}
\center
\includegraphics[trim=80 180 115 200, clip, width=0.75\textwidth]{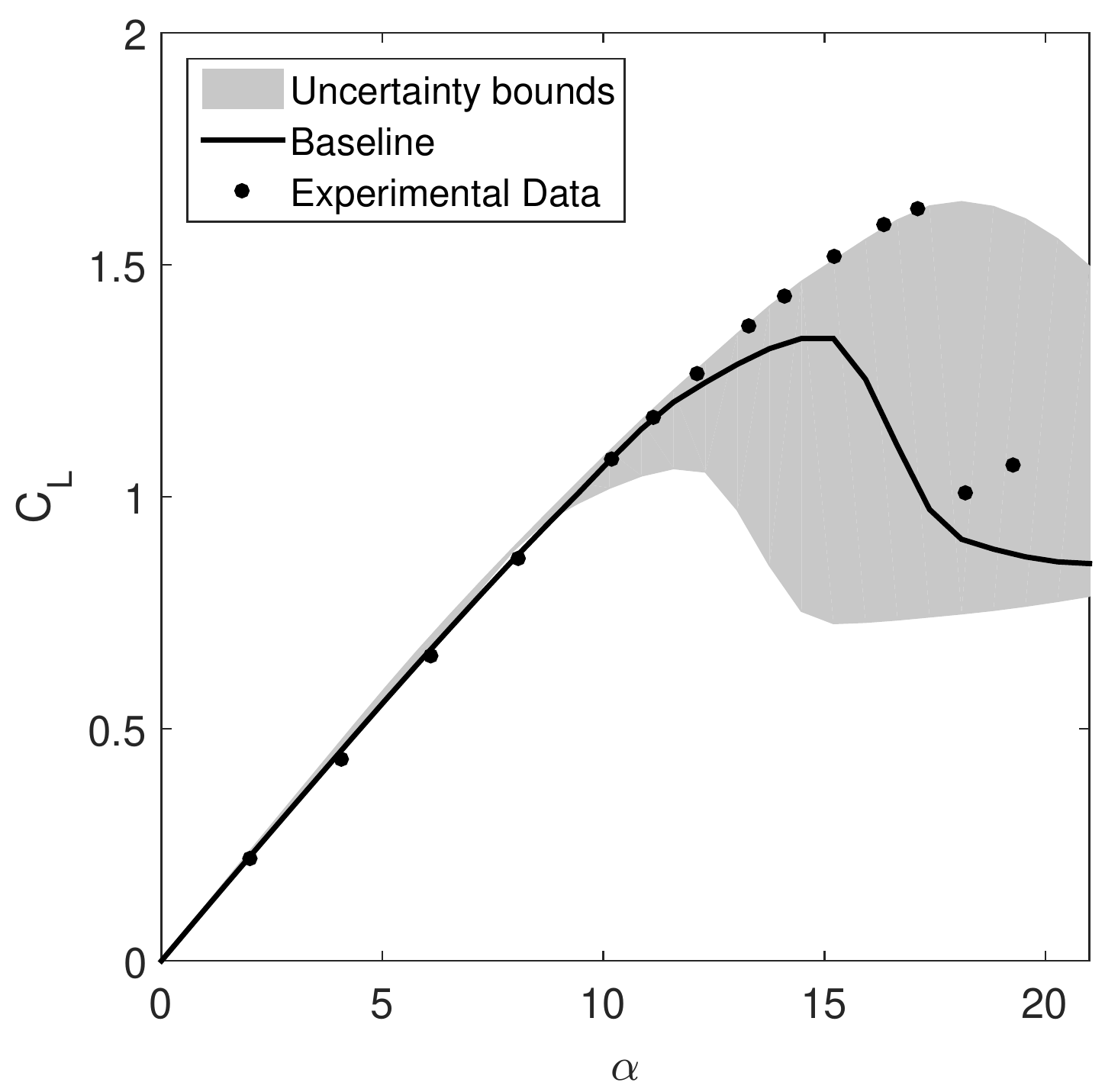}
\caption{Lift Curve of the NACA0012 with interval bounds predicted by the EQUiPS module. \label{fig:naca0012_lift_curve}}
\end{figure}

Similarly, if the module is run for a number of angles of attack, the predicted lift curve can be plotted. This showcases the robustness of the model in different flow situations. Fig. \ref{fig:naca0012_lift_curve} illustrates the results from a angle of attack sweep from 0 to 20 degrees. At low angles of attack, there is almost no discernible difference between the RANS predictions and the experimental data.  Accordingly, here the uncertainty bounds from the EQUiPS module are negligible.  At higher angles of attack closer to stall, there is substantial discrepancy between the RANS predictions and the high fidelity data. For these values of the angle of attack, the uncertainty bounds are substantial as well. At all values of the angle of attack, the uncertainty bounds from the EQUiPS module envelope the experimental data

\begin{figure}
\center
\includegraphics[trim=80 180 115 200, clip, width=.75\textwidth]{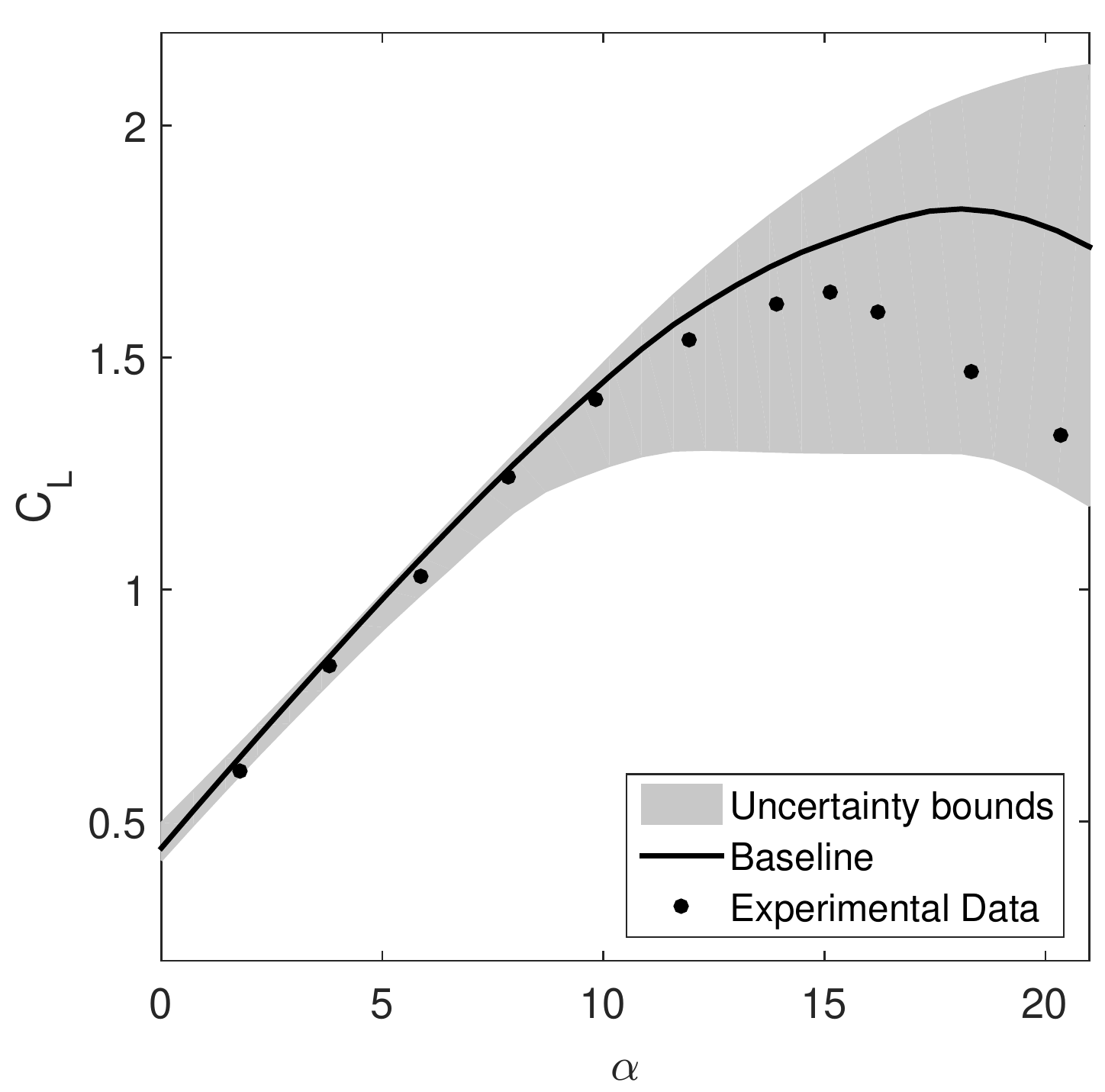}
\caption{Lift Curve of the NACA4412 with interval bounds predicted by the EQUiPS module. \label{fig:naca4412_lift_curve}}
\end{figure}

To highlight the robustness of the EQUiPS module in handling different configurations, we include a similar figure for the lift curve of the NACA4412 airfoil in Fig. \ref{fig:naca4412_lift_curve}.

\chapter{Advanced Applications} \label{3d_application}

The EQUiPS module can be used for more complex flow configurations as well. To illustrate this, the module is applied to the NASA Common Research Model (CRM) which is an aircraft geometry that was developed for applied CFD validation studies \cite{rivers_further_2012,rivers_experimental_2010}. The design is based on a Boeing 777 and is representative of a transonic commercial aircraft. It has a wealth of openly accessible computational and experimental data.

\section*{Problem Setup}

The simulation conditions are described in Table \ref{NASA_CRM_test_cond}. Note that the simulations are run at a range of angles of attack, at a free stream Mach number of 0.85. These conditions lead to complex flow features, such as shock induced separation, that  can greatly increase the uncertainties in the RANS predictions. Separated flow exists for $\alpha > 4^\circ$ at this Mach number. 
\begin{table}
\centering
    \captionsetup{justification=centering}
    \caption{Simulation conditions for the NASA CRM.} 
    \begin{tabular}{|c|c|}
        \hline
        Mach Number & $0.85$ \\ \hline
        Reynolds Number & $5\times10^6$ \\ \hline
        Reference chord length & $7.00532$ m \\ \hline
        Freestream Temperature & $310.928~\text{K}$ \\ \hline
        $\alpha$ & $-2^\circ \leq \alpha \leq 12^\circ$ \\  
        \hline
    \end{tabular}
    \label{NASA_CRM_test_cond}
\end{table}

\section*{Mesh Description}

\begin{figure}
    \centering
    \begin{subfigure}[Surface mesh of the NASA CRM.] {
        \includegraphics[trim=80 130 100 160, clip, width=.47\textwidth]{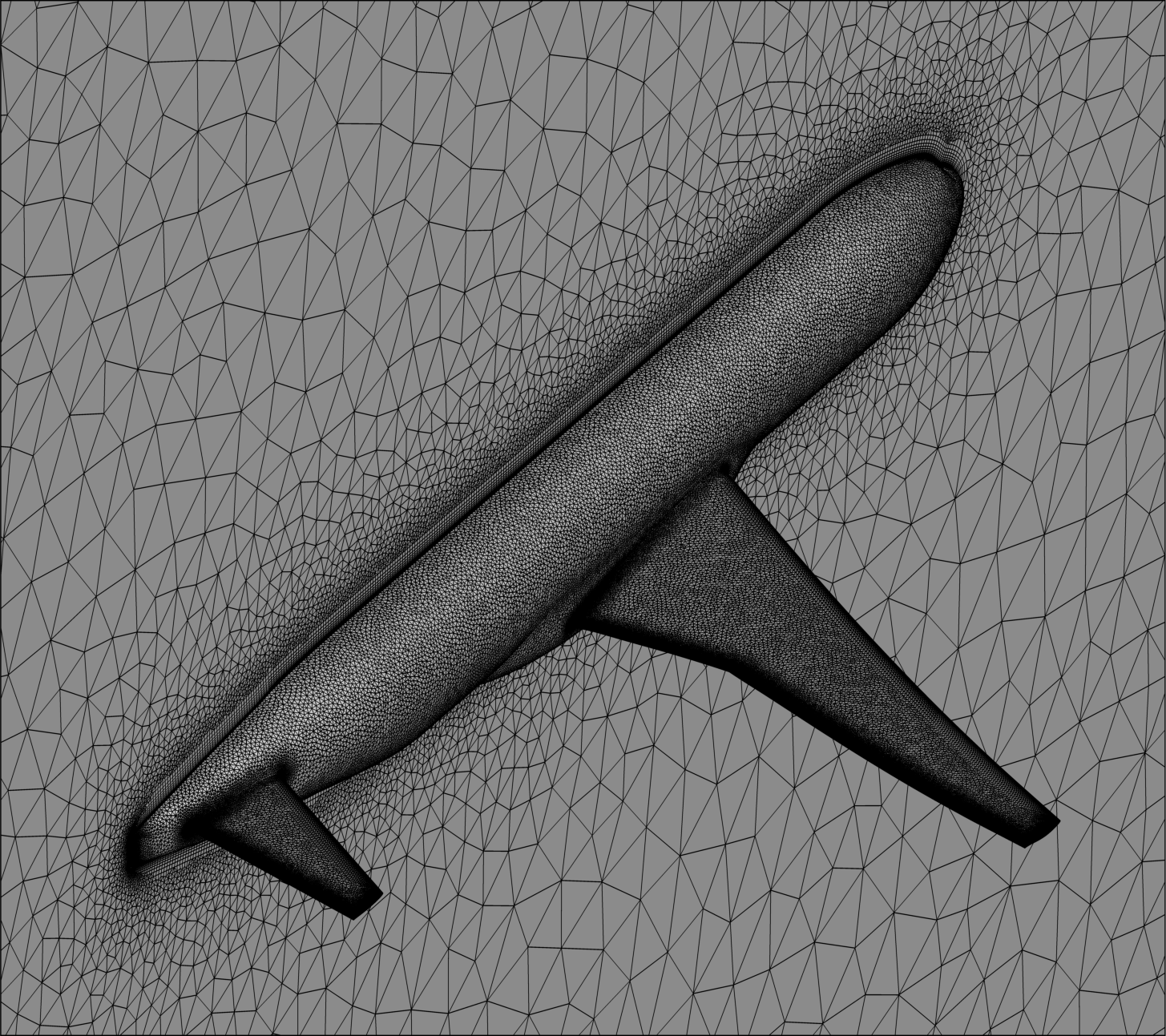} }
    \end{subfigure}
    \hfill
    \begin{subfigure}[Close up of the nose cone showing boundary layer cells on the symmetry plane.]{
        \includegraphics[trim=80 130 100 160, clip, width=.47\textwidth]{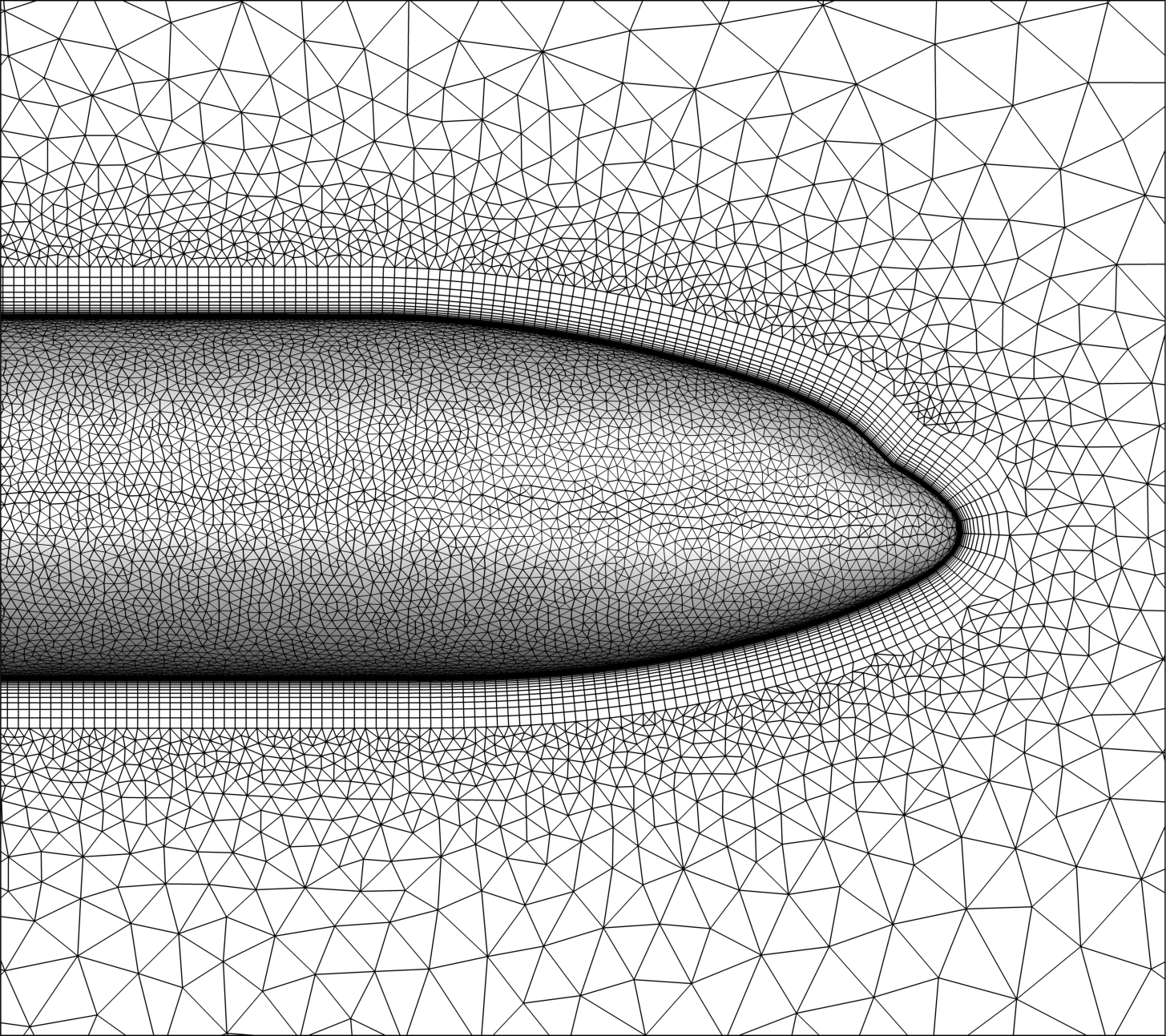} 
    }
    \end{subfigure}
    \hfill
    \begin{subfigure}[Details of the wing surface mesh.]{
        \includegraphics[trim=80 80 80 100, clip, width=.5\textwidth]{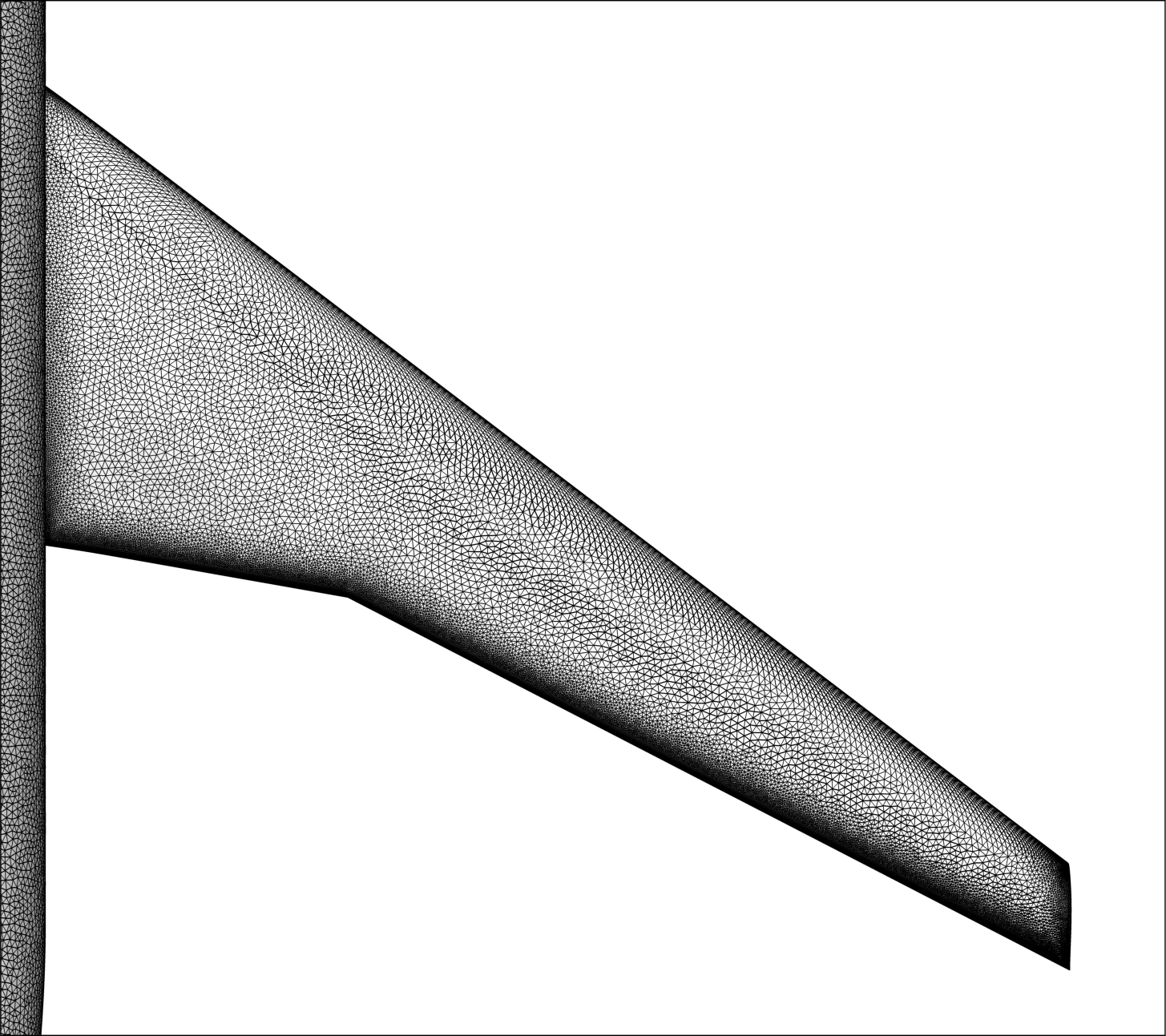} 
    }
    \end{subfigure}
    \caption{Images of the NASA CRM mesh that was used for the CFD simulations.\label{fig:crm_mesh}}
\end{figure}

Figure \ref{fig:crm_mesh} shows details of the unstructured mesh that was used for the CFD simulations. The computational domain is made of $11.8\times10^6$ mixed elements ($4.6\times10^6$ nodes) which corresponds to a coarse mesh based on the grid convergence studies performed for multiple solvers and grid topologies \cite{vassberg_summary_2010}. 

\section*{Results}

\begin{figure}
    \centering
    \begin{subfigure}[$C_L$ vs. $\alpha$.] {
        \label{fig:cl_vs_alpha}
        \includegraphics[trim=80 175 110 200, clip, width=.45\textwidth]{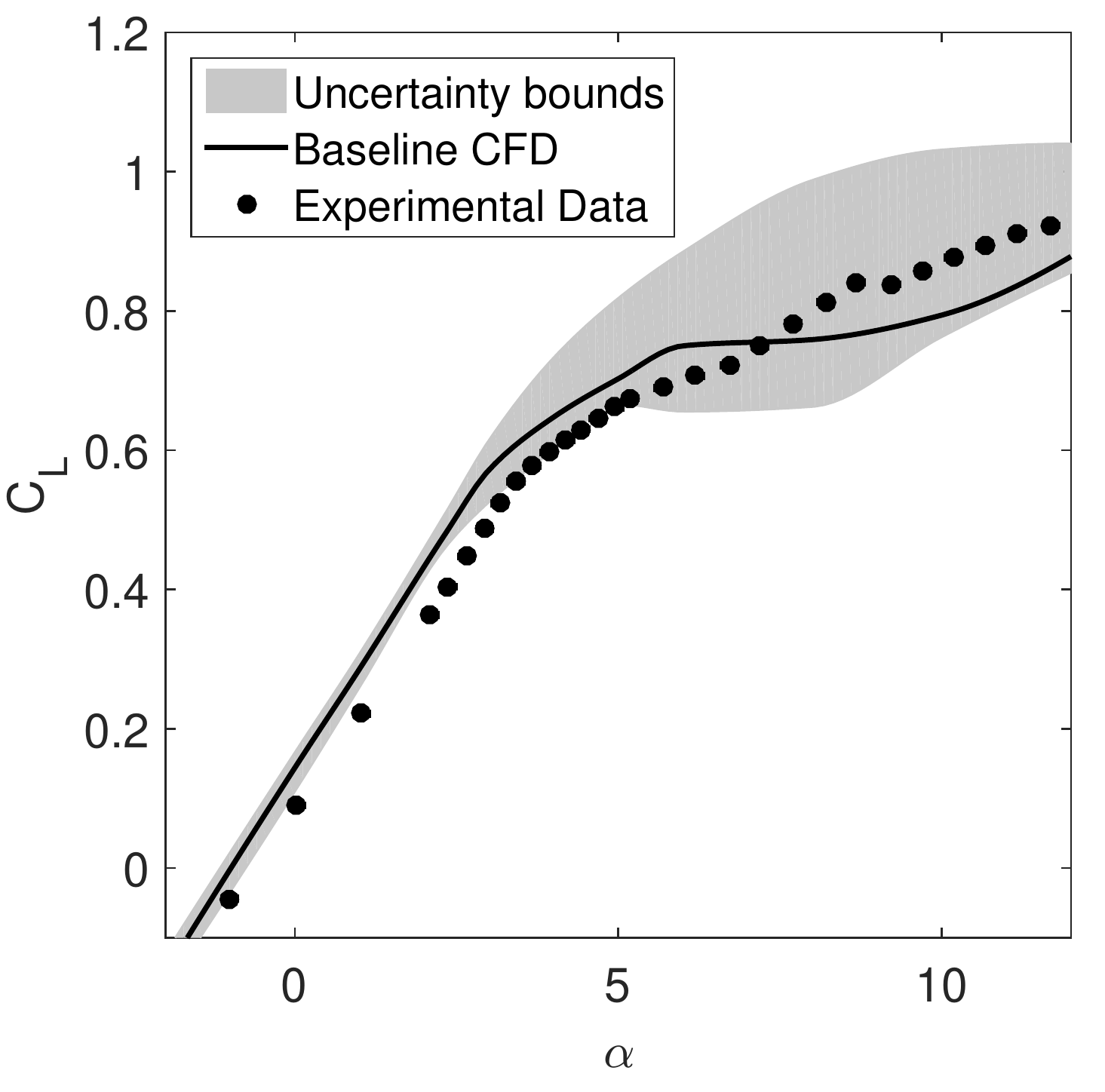}
    }
    \end{subfigure}
    \hfill
    \begin{subfigure}[$C_D$ vs. $\alpha$.]{
        \label{fig:cd_vs_alpha}
        \includegraphics[trim=70 175 110 200, clip, width=.45\textwidth]{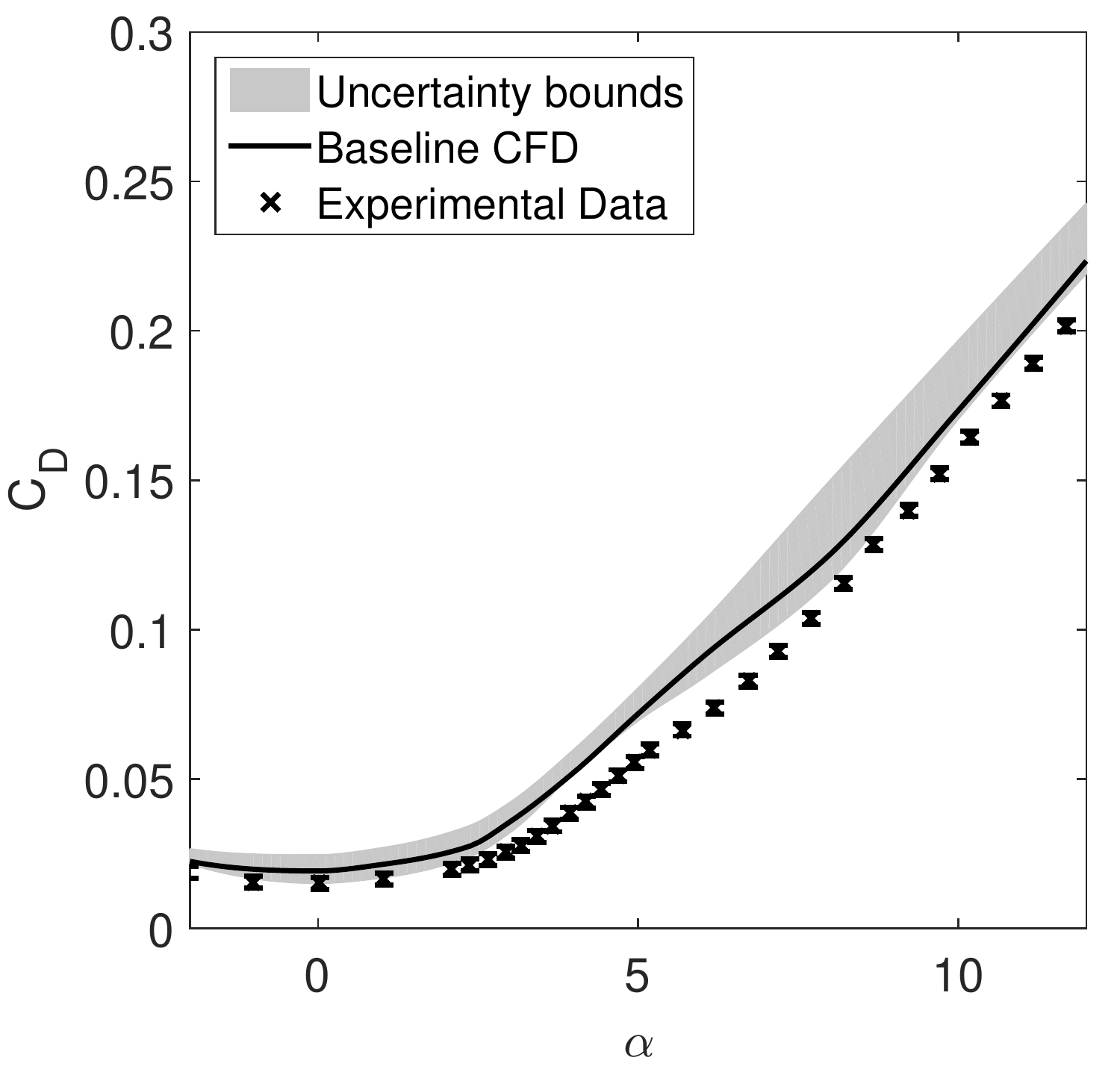} 
    }
    \end{subfigure}
    \hfill
    \begin{subfigure}[$C_m$ vs. $\alpha$.]{
        \label{fig:cm_vs_alpha}
        \includegraphics[trim=75 175 110 200, clip, width=.45\textwidth]{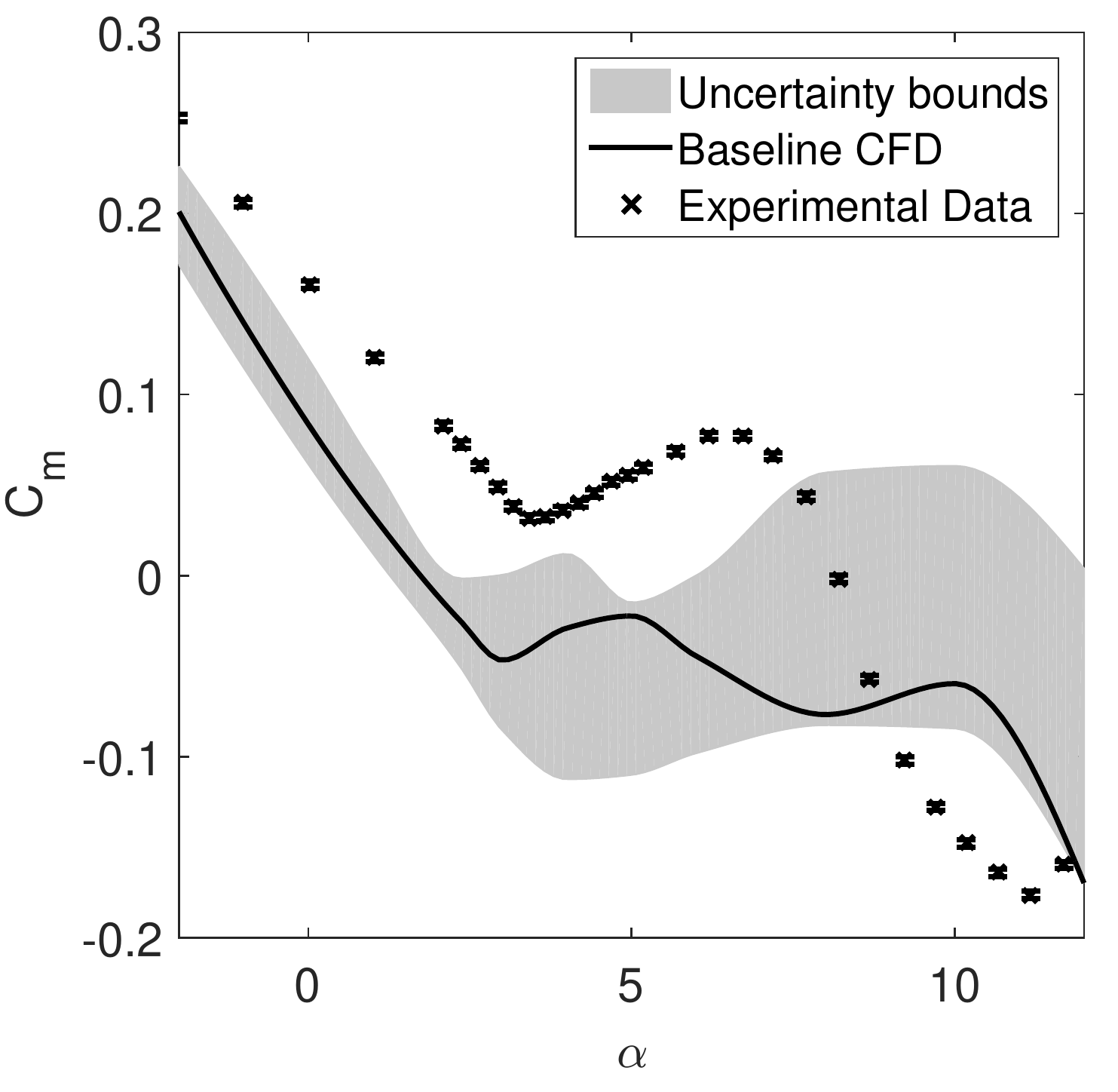} 
    }
    \end{subfigure}
    \hfill
    \begin{subfigure}[$C_L$ vs. $C_D$.]{
        \label{fig:cl_vs_cd}
        \includegraphics[trim=70 175 110 200, clip, width=.45\textwidth]{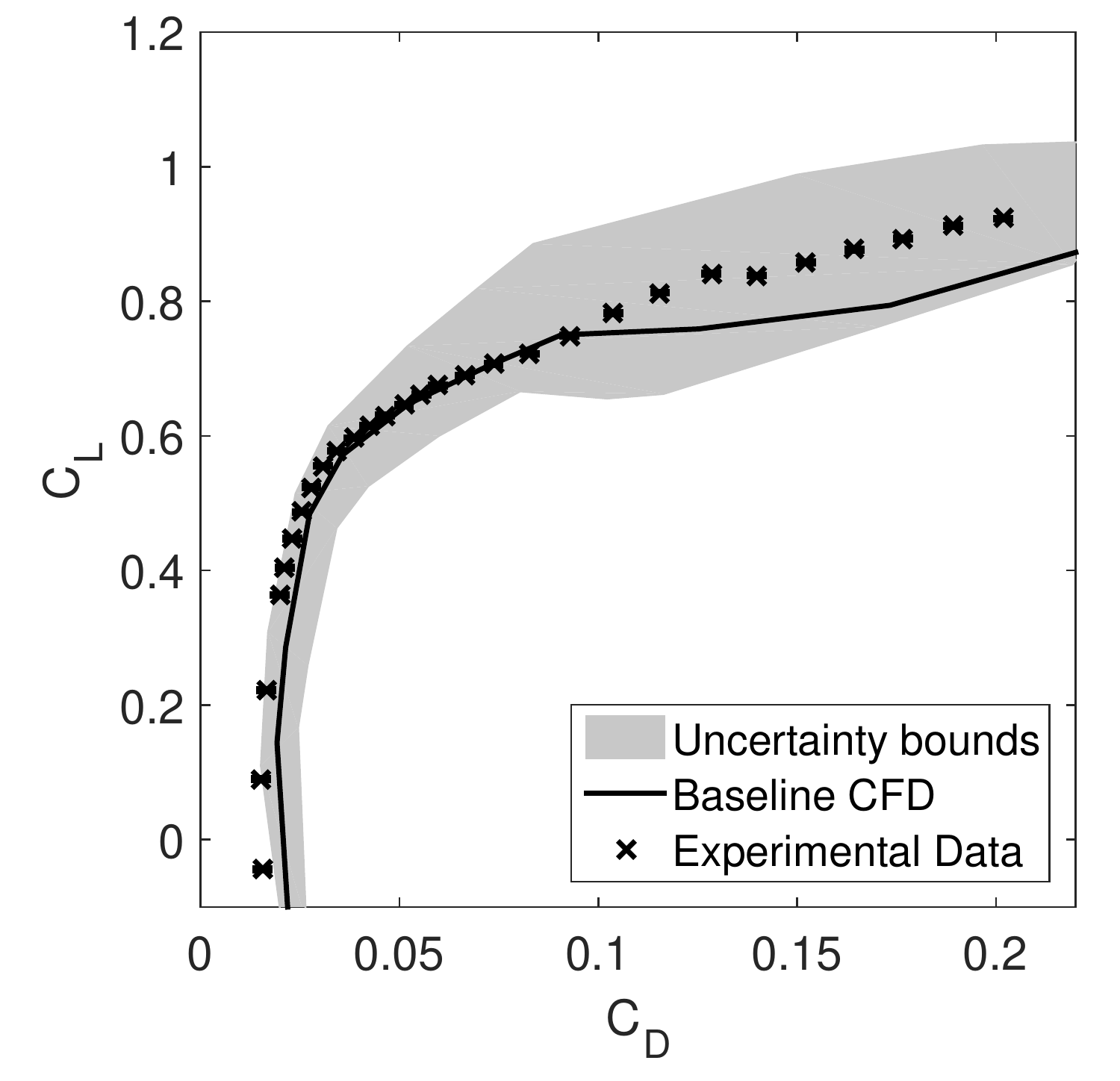} 
    }
    \end{subfigure}
    \caption{Uncertainty in force and moment coefficients as calculated by the RANS UQ methodology on the NASA CRM.\label{fig:crm_su2_uq}}
\end{figure}

The EQUiPS module is applied to the pitch sweep and compared to wind tunnel data from the NASA Ames 11ft Wind Tunnel experiment \cite{rivers_experimental_2010}. In Fig \ref{fig:crm_su2_uq}, the solid black line represents the predictions made by the baseline SST turbulence model, the grey area represents the interval bounds predicted by the EQUiPS module, and the black crosses represent the wind tunnel data. These wind tunnel data points have error bars associated with them but these are barely discernible on the scale of the plot. 

We compare the integrated quantities, specifically the coefficients of lift ($C_L$), drag ($C_D$), and longitudinal pitching moment ($C_m$), predicted by CFD and the EQUiPS module, to those experimentally determined. Focusing on the $C_L$ vs. $\alpha$ plot in Figure \ref{fig:cl_vs_alpha}. At low angles of attack, the flow remains well attached to the aircraft body and there aren't any complex flow features that would be difficult for the turbulence model to predict. The turbulence model does not introduce significant uncertainty in its predictions and, accordingly, the interval bounds predicted by the UQ module are relatively small. At higher angles of attack when there is flow separation over portions of the aircraft, simplifying assumptions made in the turbulence models make it difficult to make accurate flow predictions. This is reflected in the growing uncertainty bounds predicted by the module. This overall trend is seen in all of the plots in Figure \ref{fig:crm_su2_uq}.

Ideally, we should see better agreement between the computational and experimental data at the lower angles of attack. At these angles, the model-form uncertainty introduced by the turbulence model, as predicted by the module, is small. This discrepancy is explained by geometrical differences in the model that was experimented on, and the one that was used for simulations. The wind tunnel model of the NASA CRM underwent large aeroelastic deformations that weren't expected, and thus, not reflected in the geometry used for CFD simulations.  \cite{levy2013summary}.  

\begin{figure}
    \centering
    \begin{subfigure}[$\alpha = 1^\circ$.] {
        \label{fig:01aoa}
        \includegraphics[trim=40 300 150 280, clip, width=.45\textwidth]{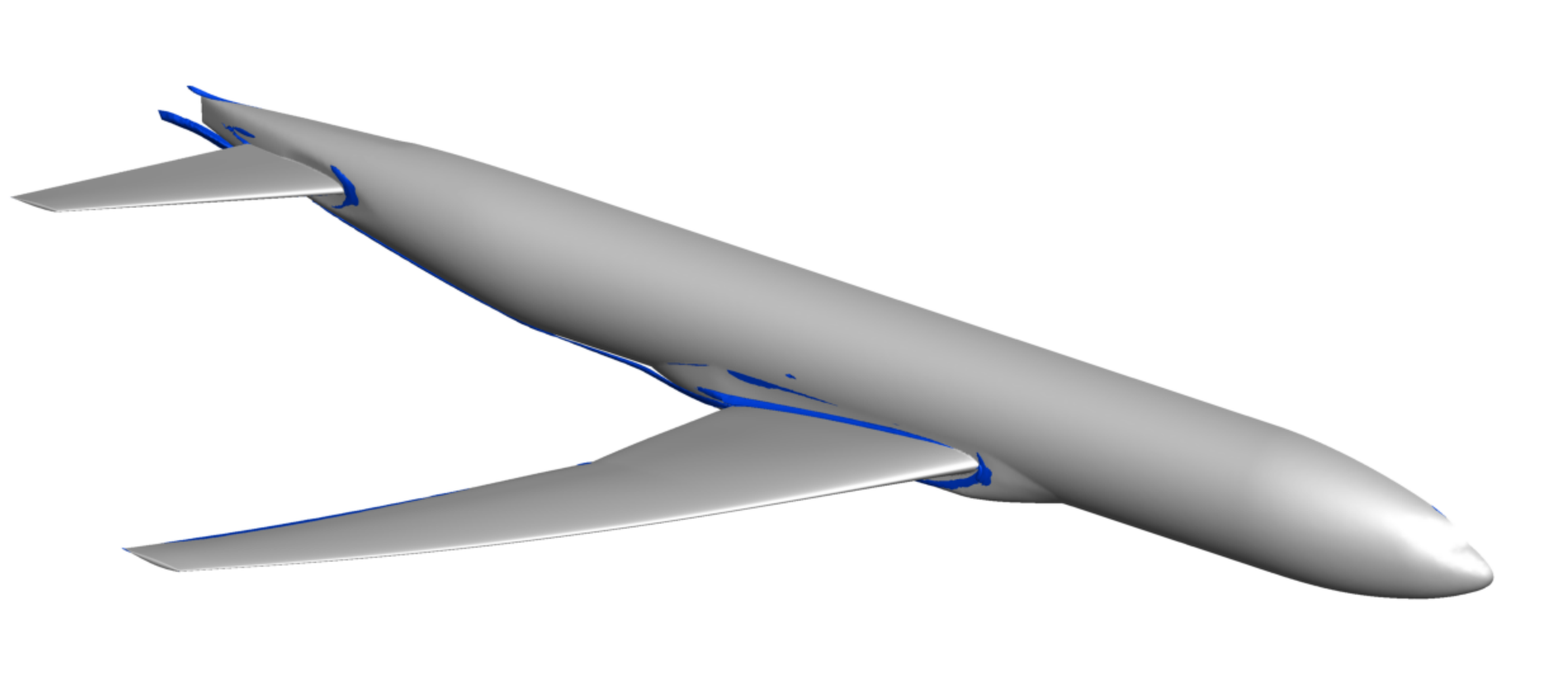} }
    \end{subfigure} 
    \hfill
    \begin{subfigure}[$\alpha = 2.35^\circ$.]{
        \label{fig:02aoa}
        \includegraphics[trim=40 300 150 280, clip, width=.45\textwidth]{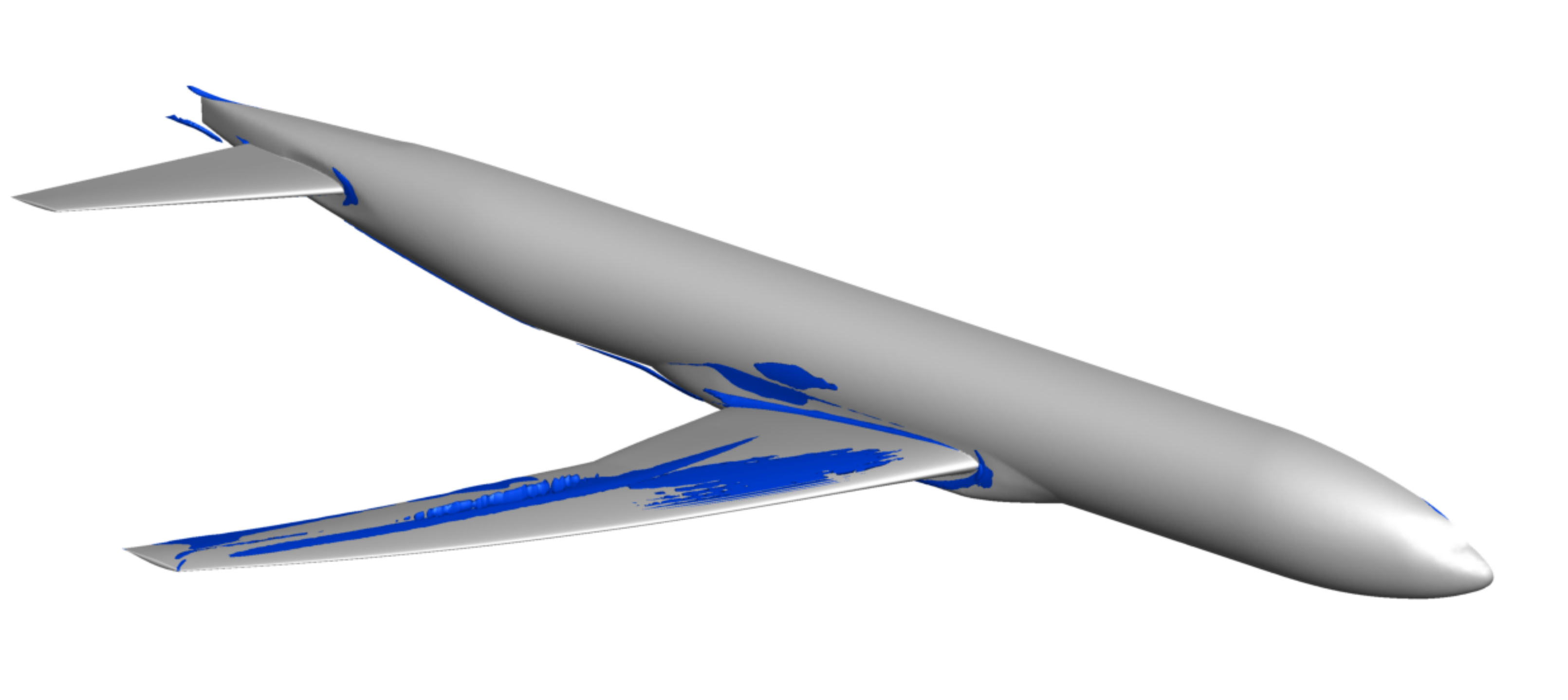} 
    }
    \end{subfigure}
    \hfill
    \begin{subfigure}[$\alpha = 3^\circ$.]{
        \label{fig:03aoa}
        \includegraphics[trim=40 300 150 280, clip, width=.45\textwidth]{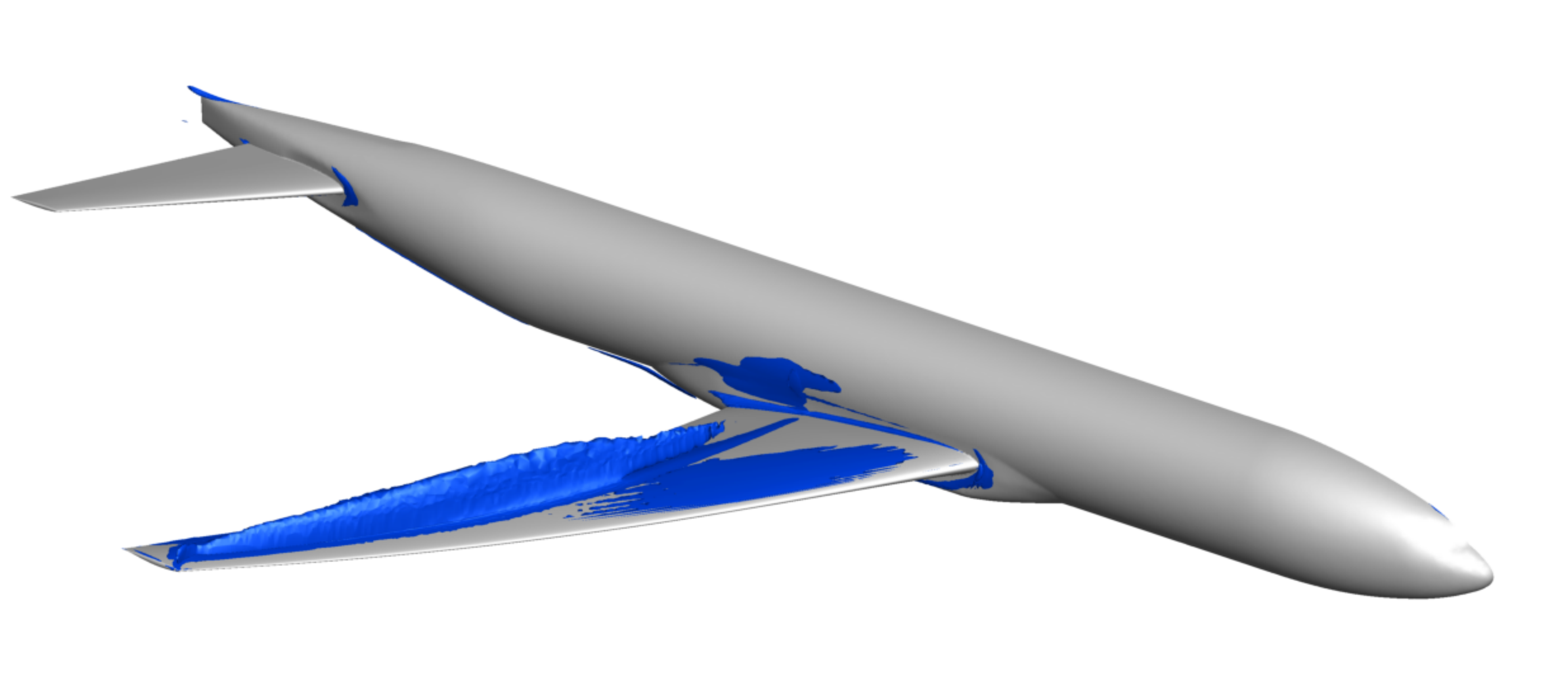} 
    }
    \end{subfigure}
    \hfill
    \begin{subfigure}[$\alpha = 4^\circ$.]{
        \label{fig:04aoa}
        \includegraphics[trim=40 300 150 280, clip, width=.45\textwidth]{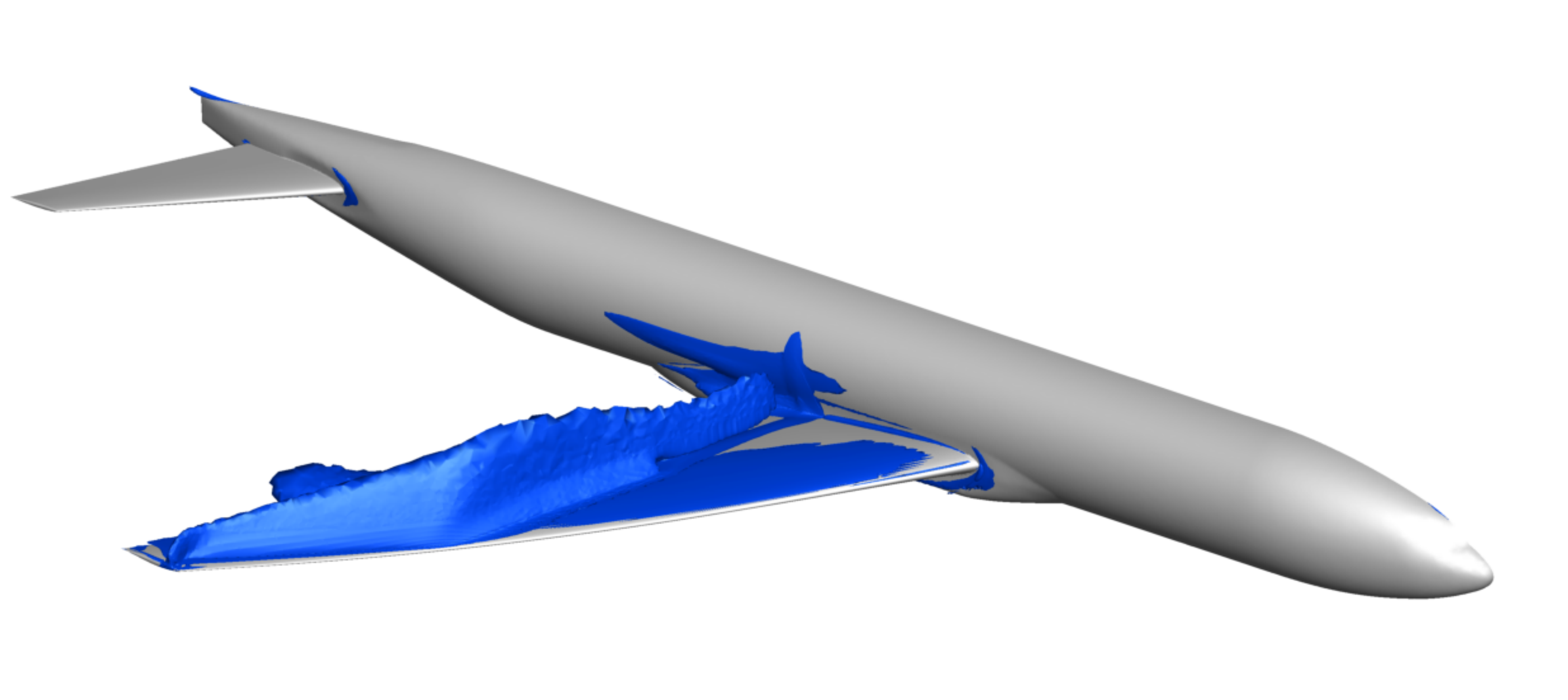} 
    }
    \end{subfigure}
    \hfill
    \caption{Isosurfaces representing areas where local Mach variability $M_v = 0.2$ at various angles of attack. \label{fig:mach_isosurface}}
\end{figure}

The EQUiPS module runs multiple RANS simulations that result in multiple realizations of the flow field. In addition to quantifying uncertainty estimates on integrated quantities, this data can provide insight into flow features/areas that contribute to the uncertainty estimates. Fig. \ref{fig:mach_isosurface} shows iso-surfaces of areas where the local Mach number varies by greater than $0.2$ across all the perturbed simulations. This Mach variability $(M_v)$ is defined at every point in the computational domain as $M_v = max(M_i) - min(M_i)$ where $i$ refers to each realization of the flow field ($5$ perturbed + $1$ baseline flow fields) and $M_i$ represents the Mach number at each point in that flow field. 

At low angles of attack, Fig. \ref{fig:01aoa}, the Mach variability is low and limited to the junction regions in the flow field. The eigenspace perturbations do not cause major changes in the flow, resulting in smaller uncertainty bounds. As the angle of attack increases, as shown in Fig. \ref{fig:02aoa} and \ref{fig:03aoa}, larger areas of variability appear where the shock would be expected, at the upper surface of the wing and away from the leading edge. This denotes an uncertainty in the shock location. This area grows rapidly until it reaches the leading edge in Fig. \ref{fig:04aoa}, signalling large uncertainty bounds and reduced confidence in the CFD predictions. Such visualizations allow us to analyse the relationship between the dominant flow features and the uncertainty that they introduce in the turbulence models. 

\chapter{Usage of EQUiPS in Published Literature} \label{usage}
Since its release, the EQUiPS module has been used extensively by researchers, and, has been acknowledged in literature. The developers of the library have used it to estimate uncertainties in RANS model predictions for benchmark flows \cite{mishra2019uncertainty} and complex flows of engineering interest \cite{mishra2020design}. Research groups from the University of Colorado, University of Michigan, Stanford University and Sandia National Laboratories used the EQUiPS module to study uncertainties in the simulation of high speed aircraft nozzles\cite{alonso2017scalable}. Additionally, researchers from the University of Greenwich have used the EQUiPS module to quantify mixed uncertainty in complex turbulent jets \cite{granados2020quantifying, granados2019influence}. Researchers from the French Institute for Research in Computer Science and Automation (INRIA) have used the EQUiPS module to carry optimization under uncertainty of Organic Rankine Cycle (ORC) supersonic nozzles \cite{razaaly2019optimization, gori2019structural}. Researchers from Stanford University along with the Boeing Company, have used the EQUiPS module to generate probabilistic aerodynamic databases \cite{mukhopadhaya2019multi}. Researchers from the University of Cambridge and the Massachusetts Institute of Technology used the EQUiPS module for design optimization under model form uncertainty for aerospace applications \cite{cook2018effective, cook2019optimization}. Researchers at the Universidad de Málaga, Spain have used the EQUiPS module for studying heat transfer characteristics in turbulent jets \cite{dfd2020a}. Similarly, investigators at the German Aerospace Center (DLR) are utilizing the EQUiPS module to investigate RANS predictions for turbomachinery applications \cite{dfd2020b}. Researchers at the University of Southampton used the EQUiPS module to study the sensitivity of aerodynamic shape optimization \cite{yang2019sensitivity}. 

In addition to these published studies, the EQUiPS module is being actively used and developed further by research groups. For instance, research groups at the Los Alamos National Laboratory are extending the EQUiPS module for the uncertainty estimation for RANS model predictions for variable-density flows. In a different vein, researchers at the Delft University of Technology are integrating data driven models to automatically tune the parameters of the perturbations in the module. We shall continue to update future versions of this document with additional research works that utilize the EQUiPS module\footnote{If you utilize the EQUiPS module in your research, please cite as:\\ @article\{mishra2019uncertainty,\\
  title=\{Uncertainty estimation module for turbulence model predictions in SU2\},\\
  author=\{Mishra, Aashwin Ananda and Mukhopadhaya, Jayant and Iaccarino, Gianluca and Alonso, Juan\},\\
  journal=\{AIAA Journal\},\\
  volume=\{57\},\\
  number=\{3\},\\
  pages=\{1066--1077\},\\
  year=\{2019\},\\
  publisher=\{American Institute of Aeronautics and Astronautics\}\}}.

\bibliographystyle{amsplain}
\bibliography{main}
\end{document}